%% file: FermionsUneqTime.tex
\documentclass[5p,twocolumn]{elsarticle}
\usepackage{mathrsfs}
\usepackage{amssymb}
\usepackage{amsmath}
\usepackage{graphicx}
\usepackage[utf8]{inputenc}
\usepackage{amsfonts,amsbsy}
\usepackage{nicefrac}
\usepackage{xcolor}
\definecolor{lcolor}{rgb}{0.5,0,0}
\definecolor{citcolor}{rgb}{0,0.3,0.0}
\usepackage[breaklinks,colorlinks,urlcolor=blue,citecolor=citcolor,linkcolor=lcolor]{hyperref}
\usepackage{comment}

\usepackage{slashed}
\usepackage{bbold}

\newcommand{\mbf}{\mathbf}

\newcommand{\as}{\alpha_\textrm{s}}

\newcommand{\tpert}{t}

\newcommand{\mrm}{\mathrm}
\newcommand{\ah}{_\mrm{ah}}

\newcommand{\HTL}{\mrm{HTL}}

\newcommand{\Q}{Q}

\newcommand{\fig}{Fig.~}

\newcommand{\eq}{Eq.~}
\newcommand{\se}{Sec.~}

\newcommand{\re}{Ref.~}
\newcommand{\res}{Refs.~}

\newcommand{\ud}{\mathrm{d}}
\newcommand{\ii}{{\boldsymbol{\hat{\i}}}}
\newcommand{\jj}{{\boldsymbol{\hat{\j}}}}

\newcommand{\pToFigs}{./}

\definecolor{azure}{rgb}{0.0, 0.5, 1.0}

\newcommand{\changeflag}[1]{{ #1}}

\begin{document}

\title{Spectral function of fermions in a highly occupied non-Abelian plasma}

\author[1]{K.~Boguslavski} 
\author[2,3]{T.~Lappi} 
\author[2,3]{M.~Mace} 
\author[4]{S.~Schlichting} 


\affiliation[1]
{organization={Institute  for  Theoretical  Physics,  Technische  Universit\"{a}t  Wien},
addressline={Wiedner Hauptstr.\ 8-10},
postcode={1040},
city={Vienna},
country={Austria}}

\affiliation[2]
{organization={Department of Physics, University of Jyv\"{a}skyl\"{a}},
addressline={P.O.~Box 35},
postcode={40014},
city={University of Jyv\"{a}skyl\"{a}},
country={Finland}}

\affiliation[3]
{organization={Helsinki Institute of Physics},
addressline={P.O.~Box 64},
postcode={00014},
city={University of Helsinki},
country={Finland}}

\affiliation[4]
{organization={Fakult\"{a}t f\"{u}r Physik, Universit\"{a}t Bielefeld},
postcode={33615},
city={Bielefeld},
country={Germany}}






\begin{abstract}
 We develop a method to obtain fermion spectral functions non-perturbatively in a non-Abelian gauge theory with high occupation numbers of gauge fields. After recovering the free field case, we extract the spectral function of fermions in a highly occupied non-Abelian plasma close to its non-thermal fixed point, i.e., in a self-similar regime of the non-equilibrium dynamics. 
 We find good agreement with hard loop perturbation theory for medium-induced masses, dispersion relations and quasiparticle residues. We also extract the full momentum dependence of the damping rate of the collective excitations. 
\end{abstract}


\maketitle

\input{Intro.tex}

\input{Method.tex}

\input{Results.tex}
\input{Conclusion.tex}


\appendix

\input{HTL.tex}

\bibliographystyle{JHEP-2modlong}
\bibliography{spires}

\end{document}

%% file: Intro.tex

\section{Introduction}

Nonperturbatively strong color fields play an important part in the physics of ultrarelativistic heavy ion collisions and the early universe. The pre-equilibrium initial stages of the collision are characterized by highly occupied glasma field configurations~\cite{Lappi:2006fp,Gelis:2015gza,Schlichting:2019abc,Berges:2020fwq} resulting from the collision of two dense gluonic systems. 
Similarly states with large occupation numbers of scalars or gauge bosons can emerge from instabilities in the reheating of the early universe \cite{Allahverdi:2010xz,Amin:2014eta,Lozanov:2019jxc}. Also in an equilibrated quark-gluon plasma, the infrared sector is dominated by gluon fields with high occupation numbers. 

Even if the energy density of the system is dominated by bosons, it is important to understand the interactions of fermions with these strong bosonic fields. In heavy-ion collisions, jets formed by light energetic quarks are created in the earliest stage of the collision. They propagate through the dense gluonic system, which has an effect on their energy loss~\cite{Ipp:2020nfu}. Electromagnetic observables provide another experimental window into the earliest stage of heavy-ion collisions~\cite{Martinez:2008di,Coquet:2021cuv}, and originate from quarks which are the only carriers of electric charge in the medium. In order to develop a microscopic description of such observables, one must understand the interactions of quarks with a system of overoccupied gluon fields. 

Since the dynamics of bosonic states with occupation numbers of the order of the inverse self-coupling $\sim 1/\as$, can be naturally described in terms of classical fields~\cite{Jeon:2004dh,Mueller:2002gd,Berges:2004yj} to leading order in $\as$, classical statistical field simulations are commonly used in heavy-ion physics~\cite{Berges:2013fga,Berges:2013eia} and cosmology~\cite{Figueroa:2020rrl}.
We will here use this classical field picture to study the microscopic properties of Dirac fermions (quarks) interacting with a over-occupied non-abelian gauge field.

The interaction of a fermion with the background field is encoded in its \emph{spectral function}. The purpose of this paper is to compute this spectral function. Based on earlier calculations of spectral functions for gluons~ \cite{Boguslavski:2018beu,Boguslavski:2021buh} and the real time lattice fermion code developed in ~\cite{Mace:2016shq,Mace:2019cqo}, we will first develop a numerical method to calculate the spectral function in an out-of-equilibrium overoccupied background gauge field configuration. We are studying fermion interactions with a strongly overoccupied gluon field, thus the dynamics is dominated by gluons and the physical situation is very different from systems at large baryon density (see, e.g., recent work in ~\cite{Tripolt:2020irx}). 
Our classical-statistical method is similar to the ones used to extract spectral functions in scalar theories both far from equilibrium \cite{Epelbaum:2011pc,PineiroOrioli:2018hst,Boguslavski:2019ecc} and for a  thermal system~\cite{Aarts:2001yx,Schlichting:2019tbr,Schweitzer:2020noq}, and can also be applied to study the dynamics of fermionic excitations in the presence of scalar or abelian gauge fields.

We will compute the spectral function in  momentum space, in both the time and frequency domains. From this spectral function we can extract medium-induced masses, dispersion relations and quasiparticle residues for the different spinor structures of the spectral function. These quantities will then be compared to predictions of hard-thermal loop (HTL) perturbation theory \cite{Braaten:1992gd,Blaizot:2001nr,Mrowczynski:2000ed,Bellac:2011kqa}. We can also extract the damping rate of fermionic quasiparticles for different momenta, which is a much more nontrivial quantity to obtain in perturbation theory~\cite{Braaten:1992gd,Baier:1991dy,Rebhan:1992ak,Blaizot:2001nr}.

This paper is structured as follows. 
We first describe the numerical method for extracting the fermion spectral function in Sec.~\ref{sec:theory}, and test it for the analytically solvable case of free fermions. We then move to a nontrivial background field in Sec.~\ref{sec:results}, where we first briefly describe the overoccupied universal UV-cascade gluon field configuration that we are using, and then present our numerical results and compare them to the expectation from HTL perturbation theory. We briefly conclude in Sec.~\ref{sec:conc}. For completeness, the HTL formulas for the spectral function from the literature are provided in \ref{sec:HTL}.

%% file: Method.tex

\section{Spectral functions from classical-statistical lattice simulations}
\label{sec:theory}

\subsection{Classical-statistical simulations}

We consider a non-abelian $\text{SU}(N_c)$ gauge theory discretized on a lattice with $N_s^3$ sites and lattice spacing $a_s$.  We use  $N_c = 2$ in this work. The gauge fields are expressed in terms of lattice gauge links $U(t',\mbf x) \approx \exp \left( i g a_s A_j(t',\mbf x) \right)$ and electric field variables $E^j(t',\mbf x) \approx g a_s^2 \partial_t A_{j}(t',\mbf x)$ in temporal axial gauge ($A_0 = 0$), where $g = \sqrt{4\pi \as}$ denotes the gauge coupling. The evolution equations for the gauge field sector then result from the lattice Hamiltonian
\begin{align}
H_{\text YM}= \frac{1}{g^2 a_s}\sum_{\mbf x,i}  \text{Tr}[E_{i}(t',\mbf x)^2] + \frac{1}{2}\sum_{j} \text{ReTr}[1-U_{ij}(t',\mbf x)],
\end{align}
where $U_{ij}(t',\mbf x) = U_i(t',\mbf x)U_j(t',\mbf x +\ii)U_i^\dagger(t',\mbf x + \jj)U_j^\dagger(t',\mbf x)$ are the usual plaquette variables and $\ii,\jj$ denote the unit lattice vectors in the $i,j=1,2,3$ spatial directions. 

While the gauge fields are treated as classical fields, fermions are described in terms of quantum mechanical field operator $\hat{\psi}(t',\mbf x)$, whose evolution is governed by the Hamiltonian
\begin{align}
 \hat{H}_W = \frac{1}{2} \sum_{\mbf x} \left[ \hat{\psi}^\dagger(t',\mbf x), \gamma^0 \left( -i\slashed{D}_s + m \right) \hat{\psi}(t',\mbf x) \right]
\end{align}
in the presence of the classical background gauge fields. With regards to the lattice discretization of fermions, we follow previous works  \cite{Mace:2016shq,Mace:2019cqo} and discretize the Hamiltonian with a tree-level improved Wilson Dirac operator
 \begin{align}
-&i\slashed{D}_s \hat{\psi}(t',\mathbf{x})= \\
 &  \frac{1}{2} \sum\limits_{n,i} C_n \Big\{  \big[-i\gamma^i-nr_W\big] U_{+n\ii}(t',\mathbf{x}) \hat{\psi}(t',\mathbf{x}+n\ii) 
\nonumber\\
&+ 2n r_W \hat{\psi}({t',\mathbf{x}}) - \big[-i\gamma^i + nr_W\big] U_{-n\ii} (t',\mathbf{x}) \hat{\psi}(t',\mathbf{x}-n\ii) \Big\}\,. \nonumber
 \end{align}
Here $i=1,2,3$ is the spatial Lorentz index and $r_W=1$ is the Wilson parameter.  The parallel transporters accross multiple lattice sites are given by products of individual link matrices, and are denoted by  $U_{+n\ii}(t',\mathbf{x})= \prod_{k=0}^{n-1} U_{i}(t',\mathbf{x}+k \ii)$ and $U_{-n\ii}(t',\mathbf{x})= \prod_{k=1}^{n} U^{\dagger}_{i}(t',\mathbf{x}-k \ii)$.
For a leading order tree-level improvement~\cite{Mace:2016shq,Mace:2019cqo} we set the coefficients $C_n$ as $C_{1}=4/3,$ $C_{2}=-1/6$ and $C_{n> 2}=0$. 

Since the equation of motion for the fermion field operator
\begin{align}
\label{eq:eom_ferm_quant}
 i \gamma^0 \partial_{0} \hat{\psi}(t',\mbf x) = \left( -i \gamma^j D_{W,j}^{s} + m \right) \hat{\psi}(t',\mbf x)
\end{align}
is linear in the fermion field $\hat{\psi}(t',\mbf x)$, it can be conveniently solved in terms of a mode function expansion~\cite{Aarts:1998td,Kasper:2014uaa}. This means that we expand the operator $\hat{\psi}(t',\mbf x)$, in terms of creation and annihilation operators of particles ($b$) and anti-particles ($d$) with definite momenta $\mbf p$ at a reference time $\tpert$
\begin{align}
 \label{eq_psi_decomposition}
 \hat{\psi}(t',\mbf x) &= \frac{1}{\sqrt{V}} \sum_{\lambda,\mbf p} \hat{b}_{\lambda,\mbf p}(\tpert)\,  \phi_{\lambda,\mbf p}^{u}(t',\mbf x) +  \hat{d}^{\dagger}_{\lambda,\mbf p}(\tpert)\,  \phi_{\lambda,\mbf p}^{v}(t',\mbf x) \, .
\end{align}
where $\lambda=1,\cdots,2N_c$ collectively labels the spin and color indices. The operator structure is determined by the action of the $\hat{b}_{\lambda,\mbf p}(\tpert)$ and $\hat{d}^{\dagger}_{\lambda,\mbf p}(\tpert)$ at a fixed reference time $\tpert$, where the creation and annihilation operators satisfy the usual equal-time anti-commutation relations
\begin{align}
\label{eq:relation_bs}
 \left\lbrace \hat{b}_{\lambda,\mbf p}(\tpert), \hat{b}^{\dagger}_{\lambda',\mbf p'}(\tpert) \right\rbrace &= 
 \left\lbrace \hat{d}_{\lambda,\mbf p}(\tpert), \hat{d}^{\dagger}_{\lambda',\mbf p'}(\tpert) \right\rbrace = V\, \delta_{\mbf p, \mbf p'}\, \delta_{\lambda \lambda'}\,.
\end{align}
With these fixed equal-time commutation relations, the time evolution of the field operator $\hat{\psi}(t',\mbf x)$ is, on the other hand, encoded in the set of $4N_c N_s^3$ wave-functions (or $N_s^3$ colored spinors)  $\phi_{\lambda,\mbf p}^{u/v}(t',\mbf x)$. They describe the propagation of a state that is given by a plane wave at the reference time $\tpert$, i.e., satisfying the initial condition
\begin{eqnarray}
\label{eq:PhiInitialCondu}
\left.\phi_{\lambda,\mbf p}^{u}(t',\mbf x)\right|_{t'=\tpert}&=&u_{\lambda}(\mbf p)e^{+ i \mbf p \cdot \mbf x}
\\
\label{eq:PhiInitialCondv}
\left.\phi_{\lambda,\mbf p}^{v}(t',\mbf x)\right|_{t'=\tpert}&=&v_{\lambda}(\mbf p)e^{-i \mbf p \cdot \mbf x}
\end{eqnarray}
Each of these wave-functions satisfies  the Dirac equation in the classical background field \eqref{eq:eom_ferm_quant}. 

Due to the large phase-space occupancy of gluons, the fermionic sector is suppressed relative to the gauge fields by one power of $\as$ in weak coupling. Working at leading order accuracy, we can therefore neglect the backreaction of fermions on the dynamical gauge fields, just as we are neglecting gluonic quantum corrections in the gluon field dynamics.  Neglecting the backreaction also makes our calculation computationally significantly less demanding, as we will explain further below. Employing a leap-frog type scheme for the discretized time evolution then results in the following update rules for the gauge field and fermion sectors~\cite{Mace:2019cqo}
\begin{align}
\label{eq:UUpdate}
& U_j(t',\mbf x) = e^{i a_t/a_s E^j(t'-a_t/2,\mbf x)}U_j(t'-a_t,\mbf x) \, , \\
\label{eq:EUpdate}
  &E^j(t'+a_t/2,\mbf x) - E^j(t'-a_t/2,\mbf x) =  \\ 
 & \qquad - \frac{a_t}{a_s} \sum_{j \neq i} \left[ U_{ij}(t',\mbf x)+ U_{i(-j)}(t',\mbf x) \right]\ah \,, \nonumber  \\
 \label{eq:PhiUpdate}
 &\phi^{u/v}_{\lambda \mbf p}(t'+a_t,\mbf x) -   \phi^{u/v}_{\lambda \mbf p}(t'-a_t,\mbf x)=\\
 &\qquad-2i a_t \gamma^{0} \left( -i \gamma^j D_{W,j}^{s}[U] + m \right)  \phi^{u/v}_{\lambda \mbf p}(t',\mbf x)\;,  \nonumber 
 \end{align}
with $[.]\ah$ denoting the anti-Hermitian part of a matrix. 
This can be solved iteratively to calculate the time evolution.%
\footnote{If not stated otherwise, we employ $a_t/a_s = 0.005$ for the discretized time evolution.}

\subsection{Spectral function of fermions}
Generally, the spectral function is defined as the expectation value of the unequal time anti-commutator of fermion fields
\begin{align}
 \rho^{\alpha \beta}(x,y) = \left\langle \left\langle \left\lbrace \hat{\psi}^{\alpha}(x), \hat{\bar{\psi}}^{\beta}(y) \right\rbrace \right\rangle_{\psi} \right\rangle 
 \label{eq_rho_def}
\end{align}
where $\hat{\bar{\psi}} = \hat{\psi}^\dagger \gamma^0$ and $\alpha, \beta$ are Lorentz indices, which should not be confused with the indices $\lambda, \lambda'$ denoting the spin and color states, that we will write as subscripts.
\changeflag{Here $\langle . \rangle_\psi$ denotes expectation values of fermionic operators in the presence of gauge fields, and $\langle . \rangle$ denotes the classical-statistical average over gauge field configurations, as is usually performed for observables in the classical-statistical framework \cite{Epelbaum:2011pc}.}
Since the spectral function has a $4\times 4$ matrix structure, it is useful to decompose it into scalar (S), pseudo-scalar (P), vector (V), axial-vector (A) and tensor (T) components according to
\begin{align}
 \label{eq_lorentz_decomp}
 \rho = \rho_S +i\gamma_5 \rho_P + \gamma_\mu \rho_V^\mu + \gamma_\mu \gamma_5 \rho_A^\mu + \frac{1}{2}\,\sigma_{\mu\nu} \rho_T^{\mu\nu}.
\end{align}
Due to rotational symmetry the vector spectral function is proportional to the momentum, and we can express its spatial components in terms of a scalar function $\rho_V $ as
\begin{align}
 \rho_V^i(\mbf p) = \frac{p^i}{E_{\mbf p}}\,\rho_V \, ,
 \label{eq:rhoV_def}
\end{align}
where $E_{\mbf p}$ is the free dispersion relation that will be discussed in Sec.~\ref{sec:free_fermions}. 
The temporal and spatial components of the spectral function can then be extracted as
\begin{eqnarray}
\rho^{0}_{V}=\frac{1}{4} \text{Tr} (\rho \gamma^{0})\;, \quad ~
\rho_{V}=- \frac{E_{\mbf p}\,p^{j}}{4\,p^2}\, \text{Tr} (\rho \gamma^{j})\;,
\end{eqnarray}
On a discrete lattice, rotational symmetry is broken, and such a relation does not hold exactly. For momentum modes far from the UV cutoff that we will consider it should, however, be satisfied. On the lattice we will determine the function $\rho_V$ by replacing $p^i$ by the effective lattice momentum $\hat{p}^i$ corresponding to the discretization of the derivative operator (Sec.~\ref{sec:free_fermions}).

By inserting the mode function expansion of the fermion fields in Eq.~(\ref{eq_psi_decomposition}) into the definition of the spectral function Eq.~(\ref{eq_rho_def}), one obtains
\begin{eqnarray}
\rho^{\alpha\beta}(x,y)= \sum_{\lambda,\mbf p} \left\langle \phi^{u,\alpha}_{\lambda,\mbf p}(x^{0},\mbf x) \left( \phi^{u,\gamma}_{\lambda,\mbf p}(y^{0},\mbf y) \right)^{*} \right. \\
\left. +\; \phi^{v,\alpha}_{\lambda,\mbf p}(x^{0},\mbf x) \left( \phi^{v,\gamma}_{\lambda,\mbf p}(y^{0},\mbf y) \right)^{*} \right\rangle \gamma_{0}^{\gamma\beta}\,, \nonumber
\end{eqnarray}
where we used the equal-time anti-commutation relations in Eq.~(\ref{eq:relation_bs}) to evaluate the anti-commutators. By changing to central and difference coordinates $\mbf X = (\mbf x + \mbf y)/2$ and $\mbf{\Delta x} = (\mbf x - \mbf y)$ in the spatial direction, we perform a spatial average over the position $\mbf X$ and a Fourier transform w.r.t.\ to difference coordinates $\mbf{\Delta x}$ according to
\begin{align}
&\rho^{\alpha \beta}(x^{0},y^{0},\mbf p) = \\
& \qquad \frac{1}{V} \int \ud^3 \mbf X \int \ud^3\Delta \mbf x~e^{-i\mbf p (\mbf x-\mbf y)} \rho^{\alpha \beta}(x^{0},\mbf x, y^{0},\mbf y)\;. \nonumber 
\end{align}
Since we are studying a system where expectation values are translationally invariant, the momentum space spectral function does not depend on the central coordinate $\mbf X$. We thus arrive at a compact expression for the spectral function in terms of the mode functions as
\begin{align}
\rho^{\alpha \beta}(x^{0},y^{0},\mbf p)
= \frac{1}{V}\sum_{\lambda,\mbf q} \left\langle \tilde{\phi}^{u,\alpha}_{\lambda,\mbf q}(x^{0},\mbf p) \left( \tilde{\phi}^{u,\gamma}_{\lambda,\mbf q}(y^{0},\mbf p) \right)^{*} \right. \\
\left. +\; \tilde{\phi}^{v,\alpha}_{\lambda,\mbf q}(x^{0},\mbf p) \left( \tilde{\phi}^{v,\gamma}_{\lambda,\mbf q}(y^{0},\mbf p) \right)^{*} \right\rangle \gamma_{0}^{\gamma\beta}\,, \nonumber
\end{align}
where $\tilde{\phi}_{\lambda,\mbf q}(x^{0},\mbf p)=\int d^3 \mbf x~\phi_{\lambda,\mbf q}(x^{0},\mbf x) e^{-i \mbf p \cdot \mbf x}$~denotes the spatial Fourier transform of the wave-functions. We note that the wave functions $\tilde{\phi}_{\lambda,\mbf q}(x^{0},\mbf p)$ depend on two momentum arguments: $\mbf q$ which is the wavenumber at the initial time $\tpert$, and $\mbf p$ which is the momentum where the wave function is evaluated. 
By choosing the reference time for the mode function expansion in Eq.~(\ref{eq_psi_decomposition}) as $\tpert=y^{0}$, we can simplify the momentum structure of the spectral function. The initial condition, Eqs.~\eqref{eq:PhiInitialCondu} and~\eqref{eq:PhiInitialCondv}, corresponds to $\tilde{\phi}_{\lambda,\mbf q}(\tpert,\mbf p) \propto \delta^{(3)}(\mbf p-\mbf q)$ in momentum space. This can be used to evaluate the sum over the momenta $\mbf q$, leading to an expression for the spectral  that is particularly convenient  for numerical evaluation 
\begin{align}
\label{eq:rho_final}
&\rho^{\alpha \beta}(x^{0},y^{0},\mbf p)= \\
&\frac{1}{V}\sum_{\lambda} \left\langle \tilde{\phi}^{u,\alpha}_{\lambda,\mbf p}(x^{0},\mbf p) u_{\lambda}^{\dagger,\gamma}(\mbf p) + \tilde{\phi}^{v,\alpha}_{\lambda,-\mbf p}(x^{0},\mbf p) v_{\lambda}^{\dagger,\gamma}(-\mbf p) \right\rangle \gamma_{0}^{\gamma\beta}. \nonumber
\end{align}

In general, the knowledge of the full set of $4N_c N_s^3$ wave-functions is required to construct the time evolution of the fermion field operator $\hat{\psi}(t',\mbf x)$ according to Eq.~\eqref{eq_psi_decomposition}. However, the spectral function in Eq.~(\ref{eq:rho_final}) can be expressed in terms of the $4N_c$ components of a single momentum mode $\phi_{\lambda,\mbf p}^{u/v}(t',\mbf x)$.
Since each momentum mode $\mbf p$ can be computed completely independently,
the calculation of the fermion spectral function is computationally significantly less demanding than simulations including the backreaction of dynamical fermions. This makes it possible to use significantly larger lattices, leading to a better resolution of different momentum scales.

Our algorithm to calculate the fermion spectral function can be summarized as follows:
\begin{enumerate}
 \item Generate a configuration of lattice gauge links $U$ and electric fields $E$, and evolve it by the classical Yang Mills equations (\ref{eq:UUpdate},\ref{eq:EUpdate}) up to the reference time $\tpert$ from which the spectral function is measured.
 \item Select a subset of $N_{\text{modes}}$ momentum $(\mbf p)$ modes, for which the spectral function is computed, and initialize the $N_\phi = 4N_c \times N_{\text{modes}}$ fermion wave-functions $\phi^{u/v}_{\lambda,\mbf p}(\tpert)$ at the reference time $\tpert$ according to Eqs.~\eqref{eq:PhiInitialCondu},~\eqref{eq:PhiInitialCondv}.
 \item Solve the Dirac equation in (\ref{eq:PhiUpdate}) for all $N_\phi$ modes along with the classical Yang Mills equations (\ref{eq:UUpdate},\ref{eq:EUpdate}) to compute the time evolution for $t'>\tpert$.
 \item Calculate the spectral function $\rho(t',\tpert,\mbf p)$ for the $N_{\text{modes}}$ momenta $\mbf p$ and $t'>\tpert$ by projecting out the appropriate plane wave component according to Eq.~(\ref{eq:rho_final}).
\end{enumerate}
This algorithm is completely analogous to the one for the gluon spectral function developed in \cite{Boguslavski:2018beu,Boguslavski:2021buh}. One initializes a fluctuation in a specific momentum mode, evolves forward in time in coordinate space, and projects back to the momentum state after the evolution. In the case of gluons, this projection involves a projection to the appropriate polarization state, while for fermions one uses the free spinors $u,v$ to project out the appropriate helicity and positive and negative energy states.

Before we proceed with the calculation of fermion spectral functions based on the above algorithm, some further comments on the gauge dependence are in order. Evidently, the fermion spectral function defined in Eq.~(\ref{eq_rho_def}) is a gauge dependent quantity, whose non-perturbative calculation requires the implementation of a suitable gauge fixing procedure. While the temporal axial gauge condition $A_0=0$ is naturally implemented in the Hamiltonian lattice gauge theory formulation, this leaves the residual gauge freedom to perform time independent gauge transformations. We eliminate the residual gauge freedom by fixing Coulomb gauge $\partial_{j} A^{j}(t,\mbf x)=0$ at the time $t'=\tpert$ when the calculation of the spectral function is initialized, i.e., between the first and the second step in the above algorithm. We note that this procedure is analogous to the linear response framework employed in \res\cite{Kurkela:2016mhu,Boguslavski:2018beu,Boguslavski:2021buh} to extract the gluon spectral function, where similarly, one fixed Coulomb gauge and subsequently studies the response of the gauge fields to plane wave perturbations in order to extract the spectral function.

\begin{figure}[tp!]
	\centering
	\includegraphics[scale=0.5]{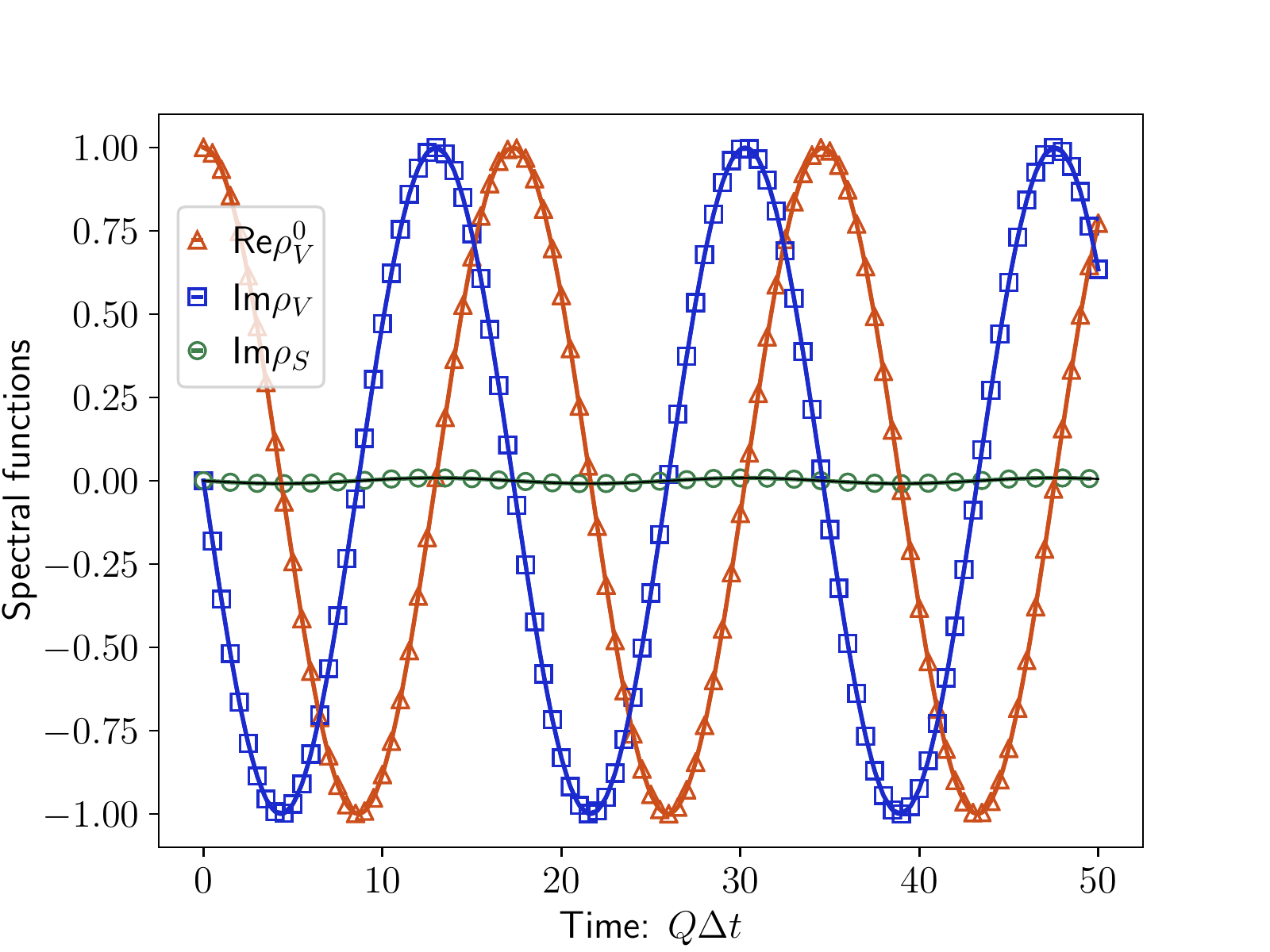}
\caption{Components of the spectral function $\rho_{S}$, $\rho_V^0$ and $\rho_V$ of free (Wilson) fermions for a fixed momentum $a_s \mbf p = (0.098, 0.195, 0.29)$ as a function of $\Delta t$. Solid curves correspond to the analytical results in \eqref{eq:free_rho}.
}
\label{fig_free_spectral}
\end{figure}

\subsection{Benchmark for free fermions}
\label{sec:free_fermions}

We first illustrate and check the method by calculating the spectral function for free fermions. This  is achieved by setting the background gauge links $U^{i}(t',\mbf x)=\mathbb{1}$ and electric fields $E^{i}(t',\mbf y)=0$. The free fermion spectral function is simply given by~\cite{Bellac:2011kqa}
\begin{equation}
  \rho^{\text{free}}(\omega,\mbf p) = 2\pi \,{\text{sgn}}(\omega)\left( p_\mu\gamma^\mu + m_{\mbf p} \right)\delta(\omega^2 - E_{\mbf p}^2) \, .
\end{equation}
In the lattice discretization the three-momentum part of the four-momentum $p^\mu$ in this expression must correspond to the discretization of the derivatives that we are using. Thus we have  $p^\mu = (\omega, \mbf{\hat{p}})$, where the effective quasi-particle momentum is 
\begin{equation}
    \hat{p}^i= -\sum_{n} \frac{C_n}{a_s} \sin\left( n\frac{ 2\pi n_i }{N_s}\right),
\end{equation} with the discrete lattice momentum mode index $n_{i}=0,\cdots,N_s-1$~\cite{Mace:2019cqo}. The Wilson term generates an effective mass that makes the doubler modes more massive (and breaks chiral symmetry), so that 
\begin{equation}
m_{\mbf p}=m + 2r_W\sum_{i,n} \frac{nC_n}{a_s}\sin^2\left( n\frac{\pi  n_i}{N_s}\right).    
\end{equation}
In terms of the effective momentum $\mbf{\hat{p}}$ and mass $m_{\mbf p}$ the energy of the single free fermion satisfies the usual relativistic dispersion relation $E_{\mbf p}=\sqrt{\mbf{\hat{p}}^2+m_{\mbf p}^2}$.
It is convenient to express the spectral function in terms of particle and anti-particle excitations as
\begin{multline}
\rho^{\text{free}}(\omega,\mbf p) \\
=  \frac{2\pi}{2E_p} \left( \Lambda_+(\mbf p)\,\delta(\omega-E_p) + \Lambda_-(-\mbf p)\,\delta(\omega+E_p) \right),
\end{multline}
where $\Lambda_{\pm}(\mbf p)$ denote the usual projections of the Dirac components%
\footnote{ Note that the particle and anti-particle projections are given by $\Lambda_{+}(\mbf p) = \dfrac{1}{N_c} \sum_\lambda u_{\lambda,\mbf p}\, \bar{u}_{\lambda,\mbf p}$ and $\Lambda_{-}(\mbf p) = \dfrac{1}{N_c} \sum_\lambda v_{\lambda,\mbf p}\, \bar{v}_{\lambda,\mbf p}$.}
\begin{align}
 \Lambda_{\pm}(\mbf p) = \gamma^0 E_{\mbf p} - \gamma^j \hat{p}^j \pm m_{\mbf p}\,.
\end{align}
The free spectral function $\rho^{\text{free}}(\Delta t,\mbf p)$ in the time domain is then obtained as
\begin{align}
 \label{eq:free_rho}
 &\rho^{\text{free}}(\Delta t,\mbf p) = \frac{1}{2E_p}\left( \Lambda_+(\mbf p)e^{-i E_p \Delta t} + \Lambda_-(-\mbf p)e^{i E_p \Delta t} \right) \nonumber \\
 &= \gamma^0 \cos(E_{\mbf p} \Delta t) + i \left( \gamma^j \frac{\hat{p}^j}{E_{\mbf p}} - \frac{m_{\mbf p}}{E_{\mbf p}} \right)\,\sin(E_{\mbf p} \Delta t)\,.
\end{align}

Comparing this to the general spinor decomposition of the spectral function in Eq.~\eqref{eq_lorentz_decomp} we see that 
the only non-vanishing components are the scalar  $\text{Im}\rho^{\text{free}}_{S}= -\frac{m_{\mbf p}}{E_{\mbf p}} \sin(E_{\mbf p} \Delta t)$, the temporal part of the vector 
$\text{Re}\rho^{0,\text{free}}_V = \cos(E_{\mbf p} \Delta t)$ and the spatial part of the vector spectral function $\text{Im}\rho^{\text{free}}_V = -\sin(E_{\mbf p} \Delta t)$. 

We show our numerical results for the free spectral function in \fig\ref{fig_free_spectral}, where we present the time evolution of the components $\text{Im}\rho_{S},\text{Re}\rho_{V}^{0}$ and $\text{Im}\rho_{V}$ calculated on a $64^3$ lattice with momentum $a_s \mbf p = (0.098, 0.195, 0.29)$ and mass parameter $m a_s = 0.003125$ corresponding to nearly massless fermions. An excellent agreement between continuous lines, depicting the analytical expressions in Eq.~\eqref{eq:free_rho}, and points, corresponding to the numerical lattice data, is observed for all components,%
\footnote{We have also checked that numerical results for the vanishing components vanish to machine precision of $10^{-16}$ for this test case.} 
validating our procedure to calculate spectral functions.

%% file: Results.tex
\begin{figure}[tp!]
	\centering
	\includegraphics[scale=0.56]{\pToFigs/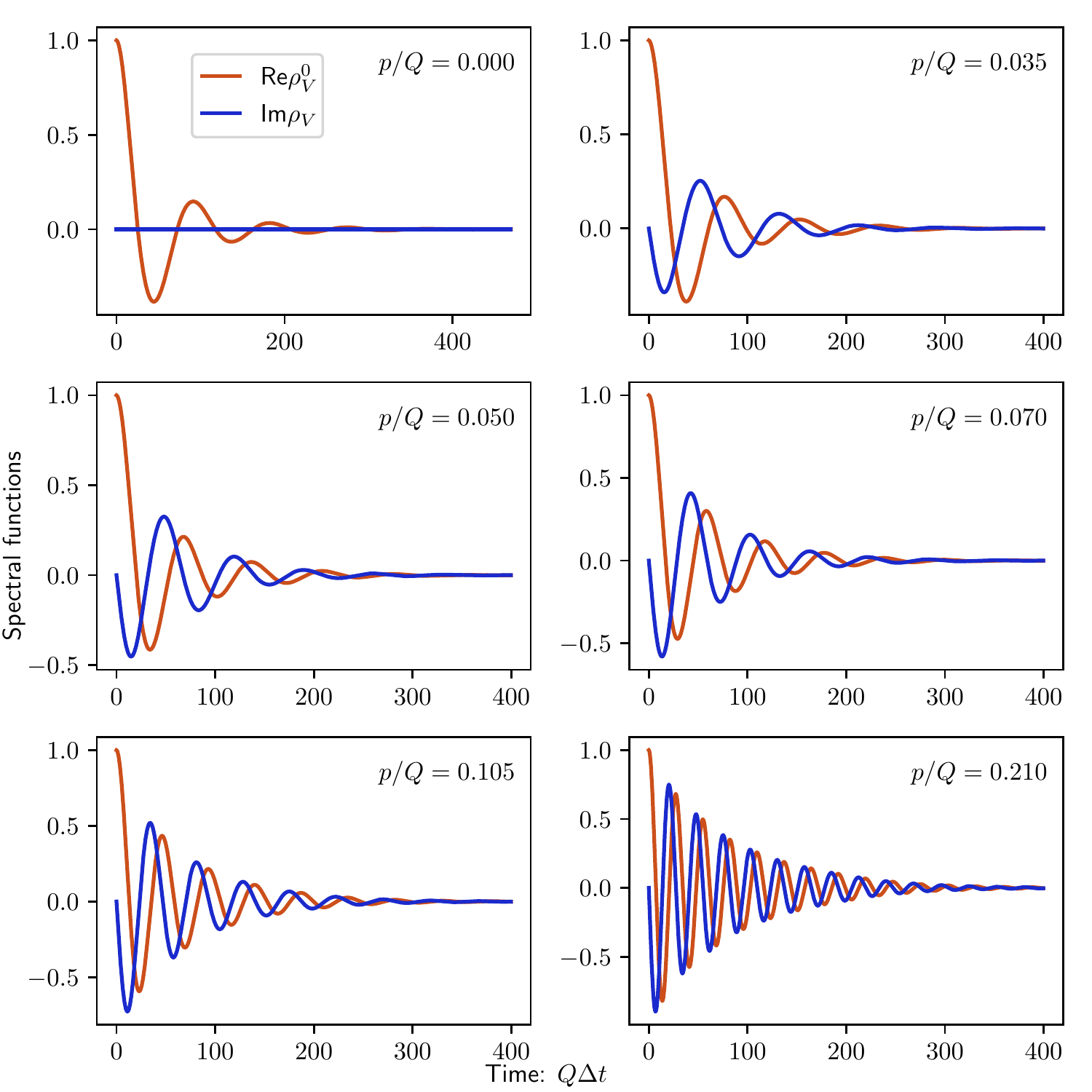}
\caption{The components of the spectral function $\rho_V^0$ and $\rho_V$ as functions of $\Delta t \geq 0$. 
}
\label{fig:p_dt}
\end{figure}


\section{Nonperturbatively computed spectral functions}
\label{sec:results}

We now turn to the investigation of quark spectral functions in a non-equilibrium plasma. We consider a highly occupied plasma of gluons, as described by the initial phase-space distribution of gluons
\begin{align}
 \label{eq:f_initial}
 g^2 f_{g}(t=0, p) = n_0\, \frac{\Q}{p}\, e^{-\frac{p^2}{2\Q^2}},
\end{align}
with $p = |\mbf p|$ and where $n_0/g^2 \gg 1$ is the initial occupancy and $Q$ is the characteristic energy scale. \changeflag{Such initial conditions can be represented by a classical-statistical ensemble of fluctuating gauge fields, which we implement numerically as in \re\cite{Boguslavski:2018beu}.}
Such overoccupied gluonic systems have been studied in several recent works~\cite{Berges:2008mr,Kurkela:2011ti,Kurkela:2012hp,Berges:2012ev,Schlichting:2012es,Berges:2013fga,York:2014wja,Mace:2016svc,Boguslavski:2018beu}; they encounter a rapid memory loss about the details of the initial conditions, and subsequently experience a self-similar scaling behavior where the dynamics of the phase-space distribution
\begin{align}
\label{eq:f_fp}
g^2f_{g}(t,p)= (Qt)^{\alpha} f_{s}\Big((Qt)^{\beta} p/Q)\Big)\;,
\end{align}
can be described in terms of a scaling function $f_{s}$ and universal scaling exponents $\alpha=-4/7$,~$\beta=-1/7$~\cite{Berges:2008mr,Kurkela:2011ti,Kurkela:2012hp,Berges:2012ev,Schlichting:2012es,Berges:2013fga,York:2014wja,Mace:2016svc}. Since the scaling behavior in Eq.~(\ref{eq:f_fp}) can be realized for a variety of different initial conditions~\cite{Micha:2004bv,Kurkela:2012hp,Berges:2013lsa,Berges:2013fga,Boguslavski:2019fsb}, this non-thermal fixed point state represents a
generic non-equilibrium state of a highly occupied plasma, and we will calculate the quark spectral function in this self-similar scaling regime. Here we will start from a moderate occupancy of $n_0=0.2$ \footnote{We note that in order to allow for a direct comparison, our initial conditions and our choice for the extraction time $\Q \tpert$ are the same as in \re\cite{Boguslavski:2018beu}, where the gluon spectral function was extracted.} and first consider quark spectral functions at a fixed reference time $Q\tpert=1500$, which is well within the self-similar regime. We will then investigate the $Q\tpert$ dependence of the spectral functions. In the scaling solution the time dependence of the hard scale and screening scale is known, and the dependence of the fermion spectral function on the reference time $\Q \tpert$ can be used to understand its structure in terms of these microscopic scales of the gluon field configurations. Note that the dependence of the spectral function on the relative time $t'-\tpert$ happens at a much shorter timescale $\sim 1/m_g$ than the dependence of the universal cascade solution on $\Q \tpert$, which is a consequence of the self-similar dynamics.
Thus measurements of the spectral function at different $\Q \tpert$ effectively study different quasi-static systems characterized by different scale separations between the hard and soft scales.
If not stated otherwise, our simulations are performed on $N_s=256$ lattices for nearly massless fermions $m = 0.003125\,\Q$  with lattice spacing $Q a_{s}=0.75$.

\subsection{Spectral functions in relative time}
Starting from the initial conditions in \changeflag{Eq.~\eqref{eq:f_initial}}, we evolve the classical Yang-Mills simulations up to the time $\Q \tpert = 1500$, where we start the calculation of the quark spectral function. Based on the algorithm presented in \se\ref{sec:theory}, we then directly obtain the different components of the spectral function $\rho(t+\Delta t,t,p)$ in the time domain. Due to the underlying symmetries, and since we consider massless fermions, we will focus on the non-vanishing vector components $\text{Re}\rho_V^0$ and $\text{Im}\rho_V$ of the spectral function.%
\footnote{Numerically, we find that the pseudoscalar, axial vector and tensor components, as well as $\text{Im}\rho_V^0$ and $\text{Re}\rho_V$, are suppressed by at least 2 orders of magnitude.}
They are depicted in \fig\ref{fig:p_dt} for a range of momenta $p/\Q = 0, \dots, 0.21$. Based on the results in \fig\ref{fig:p_dt} one observes that the spectral function in the time domain features a damped oscillatory behavior, with
\begin{align}
 \label{eq:lorD_dt}
 \text{Re}\rho_V^0(t+\Delta t,t, p) &\approx ~~e^{-\gamma(t, p) \Delta t} \cos(\omega(t, p) \Delta t)\;, \nonumber \\
 \text{Im}\rho_V(t+\Delta t,t, p) &\approx -e^{-\gamma(t, p) \Delta t} \sin(\omega( t, p) \Delta t)\,.
\end{align}
Clearly, the main differences to the free spectral function discussed in Sec.~\ref{sec:free_fermions} concern the finite damping rate $\gamma(p)$ as well as the non-trivial dispersion relation $\omega(p)$, which is nonzero even at $p=0$ due to the (non-)thermal mass induced by the medium.

\begin{figure}[tp!]
	\centering
	\includegraphics[scale=0.68]{\pToFigs/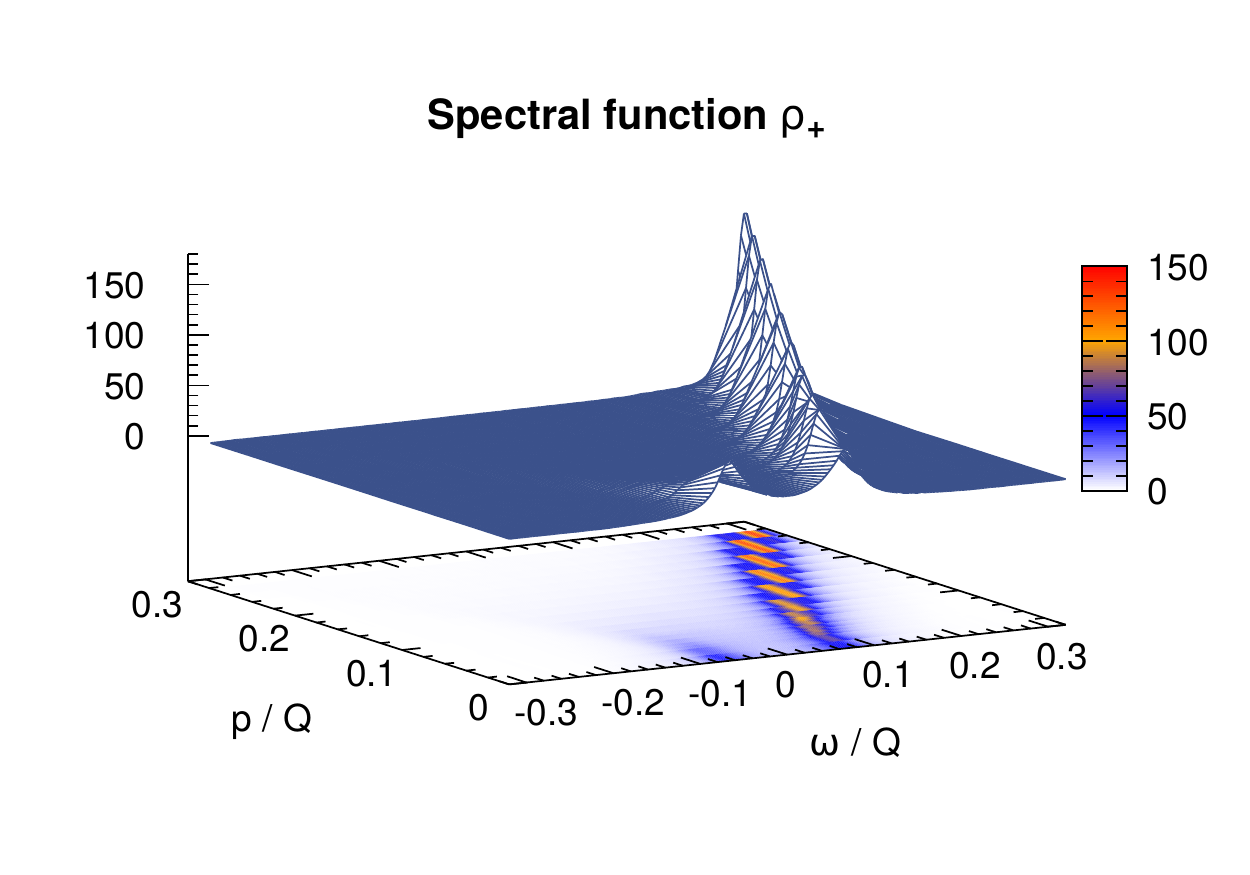}
\caption{The spectral function $\rho_+(\tpert,\omega,p)$ at $\Q \tpert = 1500$. 
}
\label{fig:3D}
\end{figure}

\begin{figure*}[tp!]
	\centering
	\includegraphics[scale=1.0]{\pToFigs/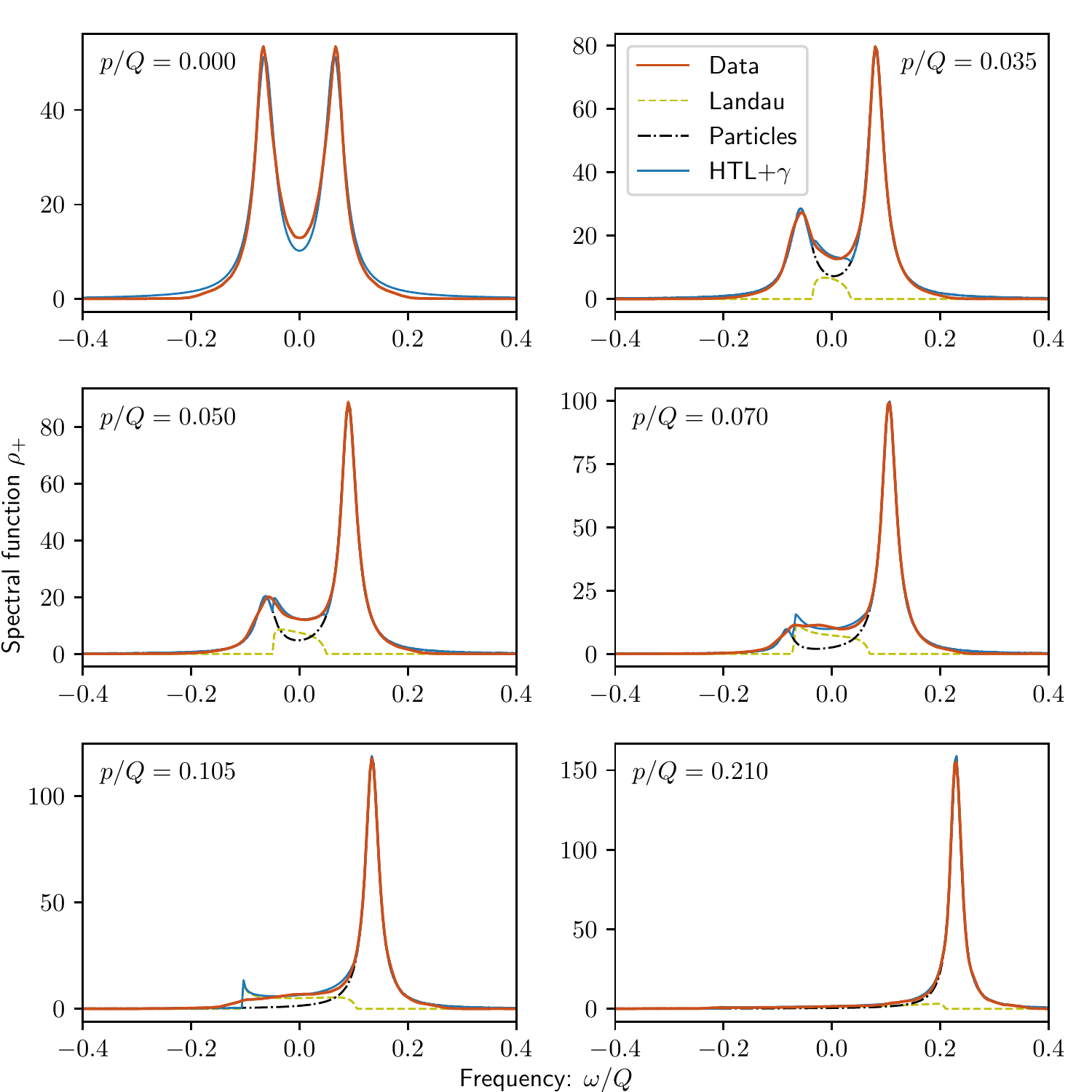}
\caption{The spectral function $\rho_+$ as a function of $\omega$. The HTL predicted Landau damping part (green dashed) and a fit to the quasiparticle peaks (black dash-dotted) are shown separately, together with the HTL curve resulting from their sum as in \eqref{eq:rhoPlus_fit} (blue continuous). As in all figures, error bars are shown for the data curves. They are computed as the standard error of the mean and are here of the order of the line width. 
}
\label{fig:p_w}
\end{figure*}

\subsection{Spectral functions in the frequency domain}
Next, in order to obtain the corresponding spectral functions in the frequency domain, we perform a Fourier transform with respect to the time difference $\Delta t = t'-\tpert$ according to
\begin{align}
\label{eq:Fourier}
 \rho_V^0(t,\omega,p) &= ~~~2\!\int_{0}^{\infty} \ud \Delta t\, \cos( \omega \Delta t)\, \text{Re}\rho_V^0\left(\tpert +\Delta t,\tpert, p\right)\;, \nonumber \\
 \rho_V(t,\omega,p) &= -2i\!\int_{0}^{\infty} \ud \Delta t\, \sin( \omega \Delta t)\, \text{Im}\rho_V\left(\tpert +\Delta t,\tpert, p\right)\;, \nonumber \\
\end{align}
where we assumed that $\text{Re}\rho_V^0\left(t+\Delta t,\tpert, p\right)$ and $\text{Im}\rho_V\left(t+\Delta t,\tpert, p\right)$ are even / odd functions in $\Delta t$ for fixed reference time $t$. We note that in practice, the integrals in Eq.~\eqref{eq:Fourier} are approximated with $Q \Delta t_{\rm max} \sim 400 - 500$ for the upper integration limit. We use zero padding, which implies that we interpret the Fourier transform as a usual integral with a continuous argument $\omega$ that we evaluate using standard integration techniques at more intermediate frequencies than provided by a discrete Fourier transform. 
\changeflag{We have checked that using a Hann windowing function in the Fourier transformation similarly to \re\cite{Boguslavski:2021buh} did not change the results.}

We provide a compact summary of our results in \fig\ref{fig:3D}, where we present a three dimensional view of the behavior of the quark spectral function 
\begin{eqnarray}
\rho_+(\tpert,\omega,p) = \rho_V^0(\tpert,\omega,p) + \rho_V(\tpert,\omega,p)
\end{eqnarray}
as a function of frequency $\omega$ and momentum $p$, noting that based on Eq.~(\ref{eq:Fourier}) the corresponding spectral function for anti-quarks $\rho_{-}(\tpert,\omega,p) = \rho_V^0(\tpert,\omega,p) - \rho_V(\tpert,\omega,p)$ can be directly obtained as $\rho_{-}(\tpert,\omega,p) = \rho_{+}(\tpert,-\omega,p)$. Starting from a symmetric spectral function at zero spatial momentum $p=0$, one observes that the spectral function becomes asymmetric along the frequency direction for $p > 0$, with a dominant peak at a positive frequency $\omega_+(p)$ and a rapidly decreasing peak at negative frequency $\omega_-(p)$. While the positive frequency peak corresponds to the usual quasi-particle excitation of a quark, the excitations at $\omega_-$ are referred to as `antiquark holes' or `plasminos' and arise from collective excitations, which emerge in thermal equilibrium \cite{Blaizot:2001nr,Bellac:2011kqa} or in a non-equilibrium state as in this work.

\begin{figure}[tp!]
	\centering
	\includegraphics[scale=0.5]{\pToFigs/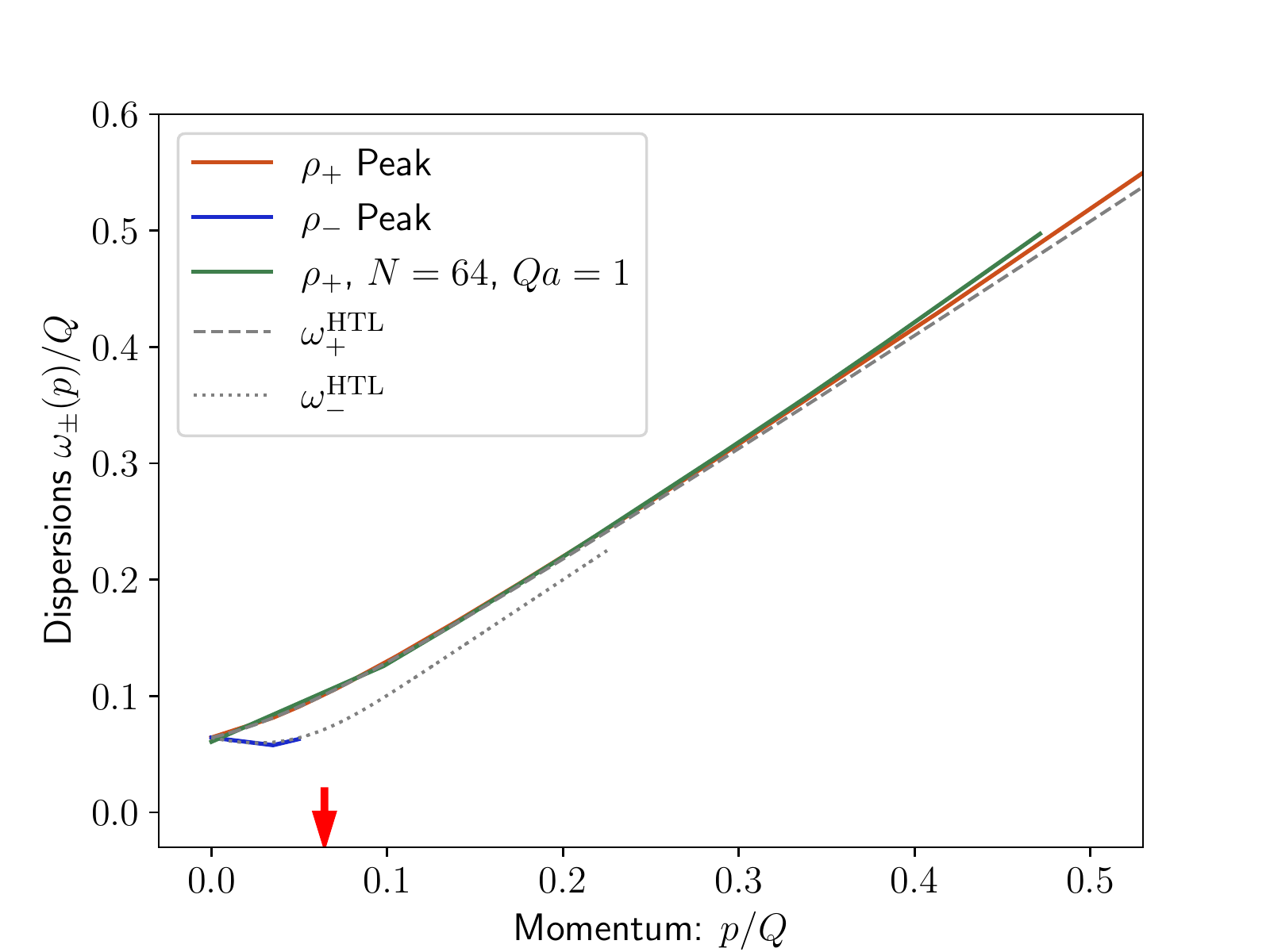}
	\includegraphics[scale=0.5]{\pToFigs/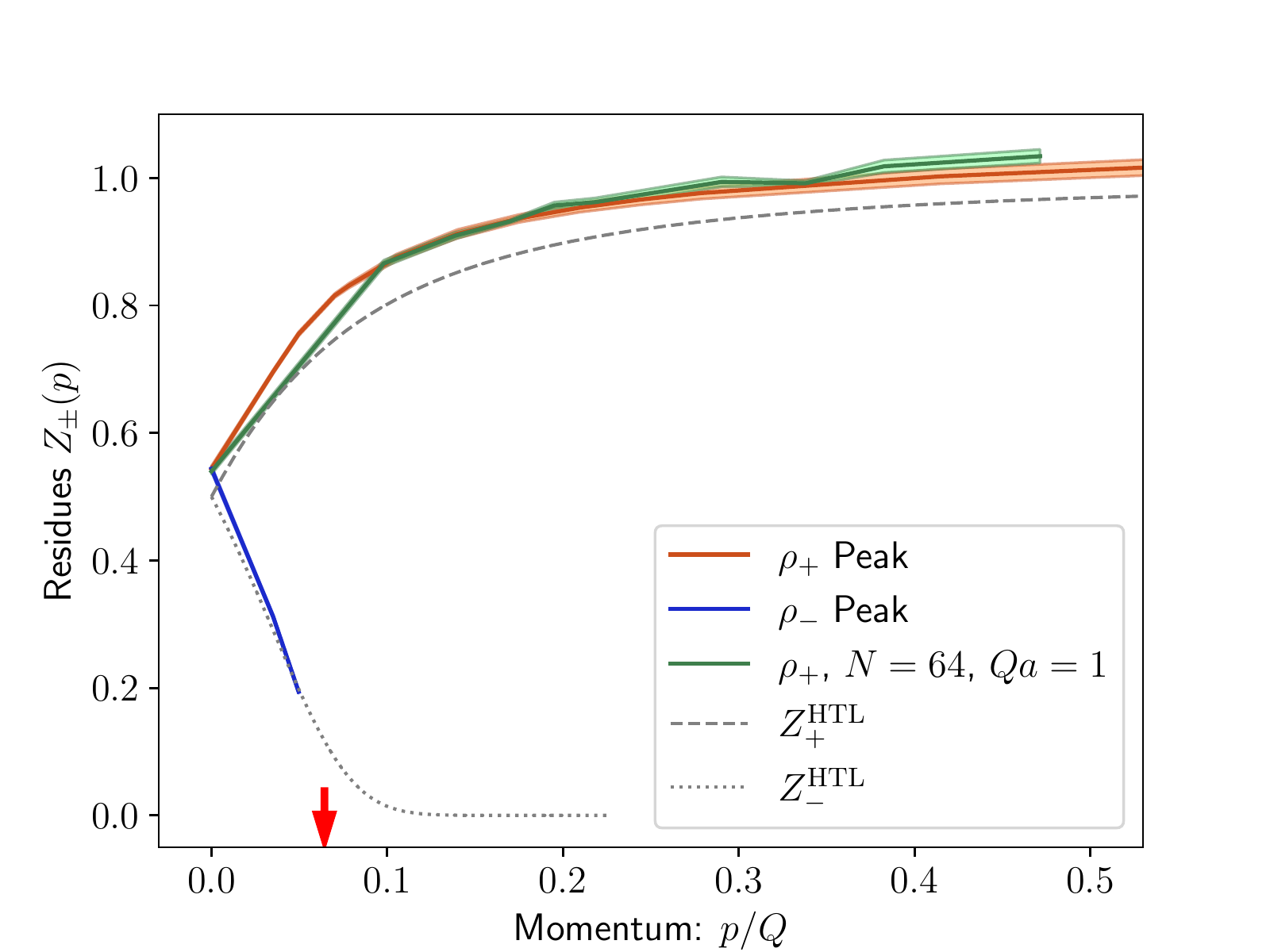}
	\includegraphics[scale=0.5]{\pToFigs/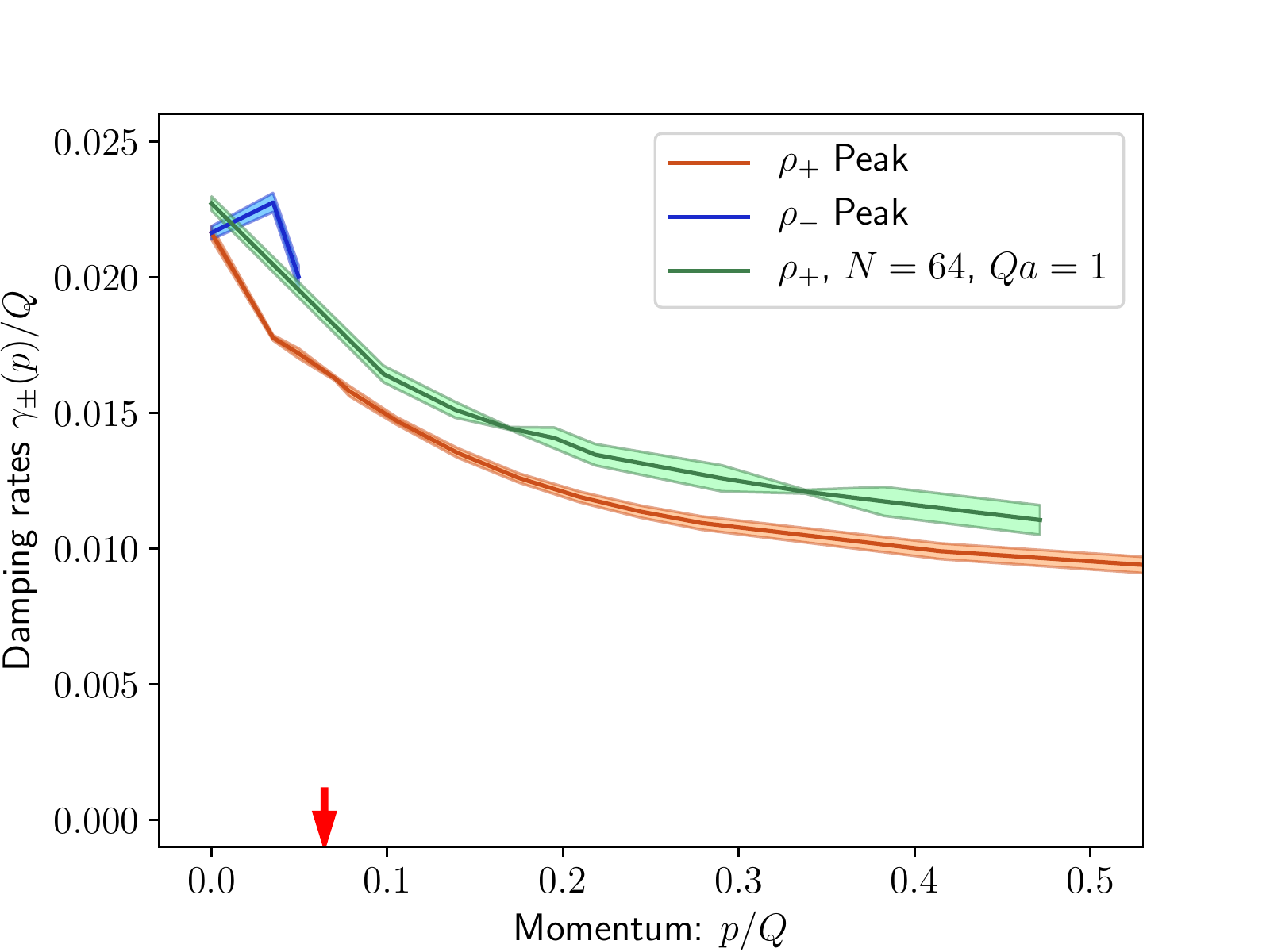}
\caption{Extracted values for the dispersion relations $\omega_\pm(p)$, residues $Z_\pm(p)$, and damping rates $\gamma_\pm(p)$ using \eqref{eq:rhoPlus_fit}. 
Results from a smaller lattice with $N=64$, $Q a_s = 1$ are shown for comparison.
HTL predictions for $\omega^\HTL_\pm(p)$ and $Z^\HTL_\pm(p)$ are added as gray dashed or dotted lines. 
The red arrows show the position of the fermion mass $m_f$. 
}
\label{fig:p_fit}
\end{figure}

\subsection{Comparison to HTL perturbation theory}

The properties of the gluon spectral function in the same field configurations that we are studying have been extensively compared to HTL perturbation theory in Ref.~\cite{Boguslavski:2018beu}. We will here perform a similar comparison for the quark spectral function.
The general structure of the HTL spectral function is given by
\begin{align} 
 \label{eq:HTL_rho}
 &\rho_{+}^{\rm HTL}(\omega, p)= 2\pi\,\beta_+(\omega/p,p) \\
 & \qquad +2\pi \left[ Z_+(p)\delta(\omega - \omega_{+}(p)) + Z_-(p)\delta(\omega + \omega_{-}(p)) \right]\;, \nonumber
\end{align}
where $\omega_{\pm}(p)$ and $Z_{\pm}(p)$ denote the positions and residues of the quasi-particle and plasmino poles, while $\beta_+$ describes the contribution from the Landau cut. We provide the detailed expressions in App.~\ref{sec:HTL}, noting that in HTL perturbation theory, these quantities are uniquely determined in terms of the momentum $p$ and the  quark screening mass $m_f$. In leading order HTL perturbation theory this is given by
\begin{align} \label{eq:mf_def}
 m_f^2 \,&= C_F \int \dfrac{\ud^3 p}{(2\pi)^3}\, \dfrac{g^2f_g(p)}{p}\;,
\end{align}
with $C_F = (N_c^2-1)/(2N_c)$. 
Within our numerical simulations we determine  the quark screening mass from the relation $m_f^2 = \dfrac{C_F}{2 N_c}\, m_{g}^2$. Here, following \cite{Boguslavski:2018beu}, the gluon asymptotic mass $m_{g}^2$ is obtained from the self-consistent solution of
\begin{eqnarray}
m_{g}^2 = \frac{2 N_c}{(N_c^2-1)} \int \frac{\ud^3 p}{(2\pi)^3}\,\frac{g^2\,\text{Tr}\left(\left\langle E_T(\mbf p) E_T^*(\mbf p) \right\rangle \right)}{p^2+m_g^2} ,
\end{eqnarray}
with the transverse field correlator $E_{T}(\mbf p) E_{T}^{*}(\mbf p) = (\delta^{ij}-p^{i}p^{j}/p^2)E_{i}(\mbf p)E_{j}^{*}(\mbf p)$. 
Once the mass parameter $m_{f}^2$ is determined, HTL perturbation theory gives us a prediction for the fermion spectral function without any further parameters. 

In order to compare our results to HTL perturbation theory, we fit our numerical data to the following functional form
\begin{align}
\label{eq:rhoPlus_fit}
&\rho_{+}^{\mrm{HTL}+\gamma}(\omega,p)= 2\pi \beta_{+}(\omega/p,p) \\
&+\,\frac{2Z_{+}(p) \gamma_{+}(p)}{(\omega-\omega_{+}(p))^2+\gamma_{+}^2(p)} + \frac{2Z_{-}(p) \gamma_{-}(p)}{(\omega+\omega_{-}(p))^2+\gamma_{-}^2(p)}\;, \nonumber
\end{align}
where `HTL+$\gamma$' refers to the HTL form supplemented with a finite width $\gamma$.
While the leading order HTL spectral function (see \ref{sec:HTL}) features stable quasiparticles represented by delta peaks, this parametrization allows for a finite width of the peaks, which are taken to have a Lorentzian form. The free parameters in the fit are the locations of the quasiparticle peaks $\omega_{\pm}(p),$ their residues $Z_{\pm}(p)$ and the widths 
$\gamma_{\pm}(p)$ for each value of the momentum $p$. The Landau cut contribution $\beta_{+}(\omega/p,p)$ is taken to be the one from HTL perturbation theory (see \ref{sec:HTL} for the explicit functional form). 

We demonstrate the quality of this fit, (henceforth denoted as `HTL+$\gamma$' referring to Eq.~\eqref{eq:rhoPlus_fit}) in  \fig\ref{fig:p_w}, where full fits are shown as blue continuous lines. The individual quasi-particle, plasmino and Landau damping contributions are also shown separately as black dash-dotted and green dashed lines. Overall, one observes an excellent agreement between our data and the HTL+$\gamma$ fits. Small deviations occur only for $p=0$, where the approximation of a width much smaller than the dispersion is not valid, and in the vicinity of $\omega \simeq -p$, where the Landau cut is smeared due to interactions.

We use the HTL+$\gamma$ fits to extract the dispersion relations $\omega_\pm(p)$, residues $Z_\pm(p)$, and damping rates $\gamma_\pm(p)$ separately for each momentum $p$ from our numerical lattice data, 
averaging over the direction $\mbf p/p$ where available. Error bars are obtained as the sum of the fitting error and the standard error of the mean.
The extracted values of $\omega_\pm(p)$, $Z_\pm(p)$ and $\gamma_\pm(p)$ are shown in \fig\ref{fig:p_fit} as functions of momentum, together with the HTL predictions for $\omega^\HTL_\pm(p)$ and $Z^\HTL_\pm(p)$ depicted as dashed or dotted lines.
Beside the extracted values for $N_s=256$, $Q a_{s}=0.75$, we also show results obtained with larger lattice spacing $N=64$, $Q a_s = 1$, to indicate that, apart from possibly the width $\gamma$, the results are not very sensitive to discretization artifacts.

We find that our results for the dispersion relations and residues agree remarkably well with the predictions from HTL perturbation theory, for which red arrows indicate the position of the fermion screening mass $m_f$ computed within HTL. Even the expected non-monotonic behavior of the $\omega_-(p)$ dispersion is clearly visible in our data, and one also observes that, as expected from HTL, the plasmino excitation gets strongly suppressed for momenta $p \gtrsim m_f$.

Clearly, the most significant deviation from leading order HTL perturbation theory is the emergence of a finite decay width of quasi-particles and plasminos $\gamma_{\pm}(p)$ depicted in the lower panel of \fig\ref{fig:p_fit}. While perturbative calculations of the fermion damping rate suffer from an infrared sensitivity to the soft gauge field propagator \cite{Blaizot:2001nr},%
\footnote{
\changeflag{The analytical expression for $\gamma(p{=}0)$ has been calculated in thermal equilibrium in Ref.~\cite{Braaten:1992gd}. 
However, this calculation does not directly give a precise estimate in the overoccupied regime. We are not aware of an extension of this calculation to our non-equilibrium system. 
It is interesting to note that in thermal equilibrium the gluon \cite{Braaten:1990it} and fermion \cite{Braaten:1992gd} damping rates are similar in magnitude whereas for our system the quark damping rate is an order of magnitude larger than the gluon damping rate extracted in Ref~\cite{Boguslavski:2018beu}.} 
}
our non-perturbative calculation can yield first principles insights into the magnitude and momentum dependence of the damping rate. Generally, we find that $\gamma_+(p)$ is smaller, but of comparable size to the quark screening mass $m_{f}$. One also observes from \fig\ref{fig:p_fit} that the fermion damping rate $\gamma_+(p)$ decreases monotonically as a function of momentum, which is qualitatively different from gluonic spectral functions in non-equilibrium overoccupied plasmas, where a monotonically increasing damping rate has been observed \cite{Boguslavski:2018beu,Boguslavski:2021buh}.

\begin{figure}[tp!]
	\centering
	\includegraphics[width=0.48\textwidth]{\pToFigs/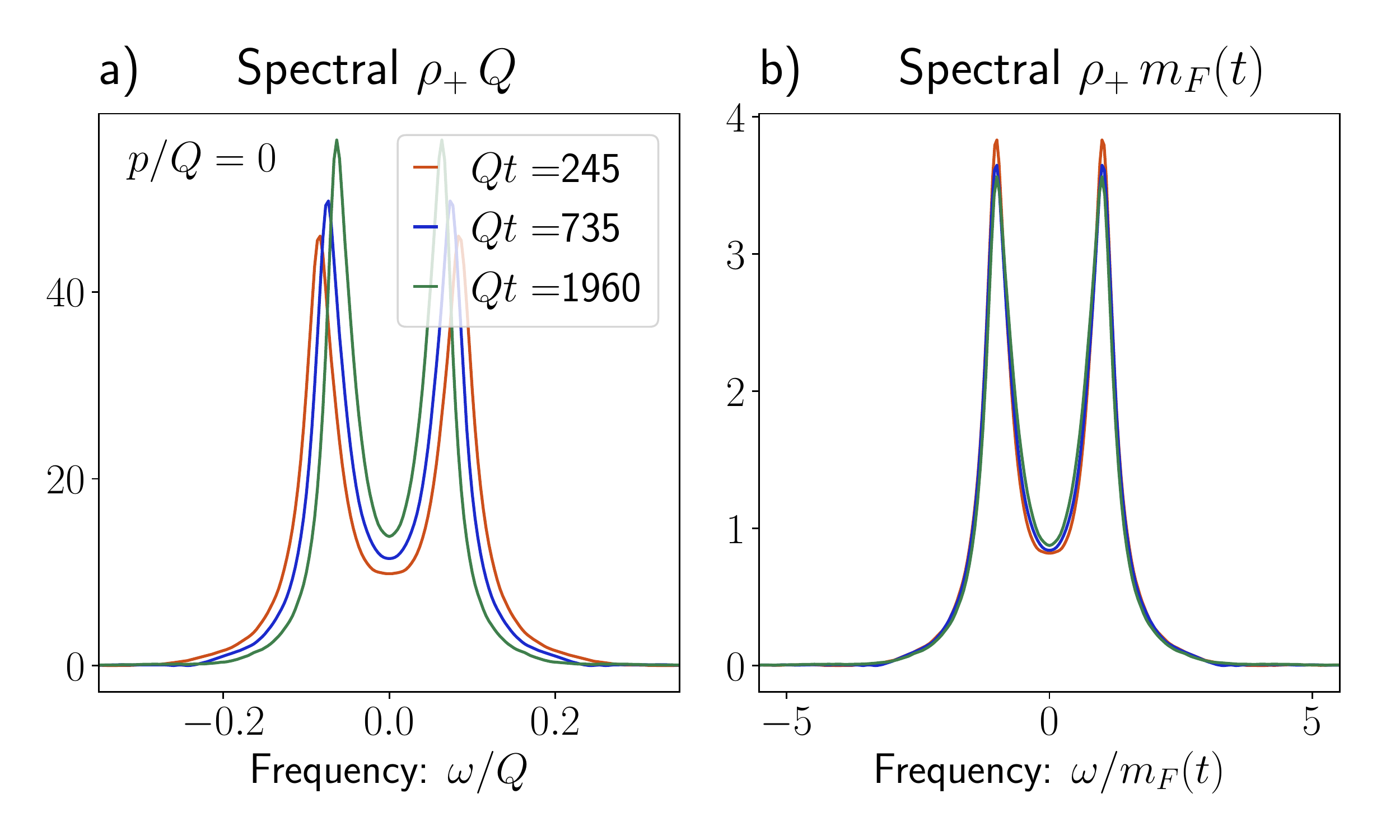}
	\includegraphics[scale=0.5]{\pToFigs/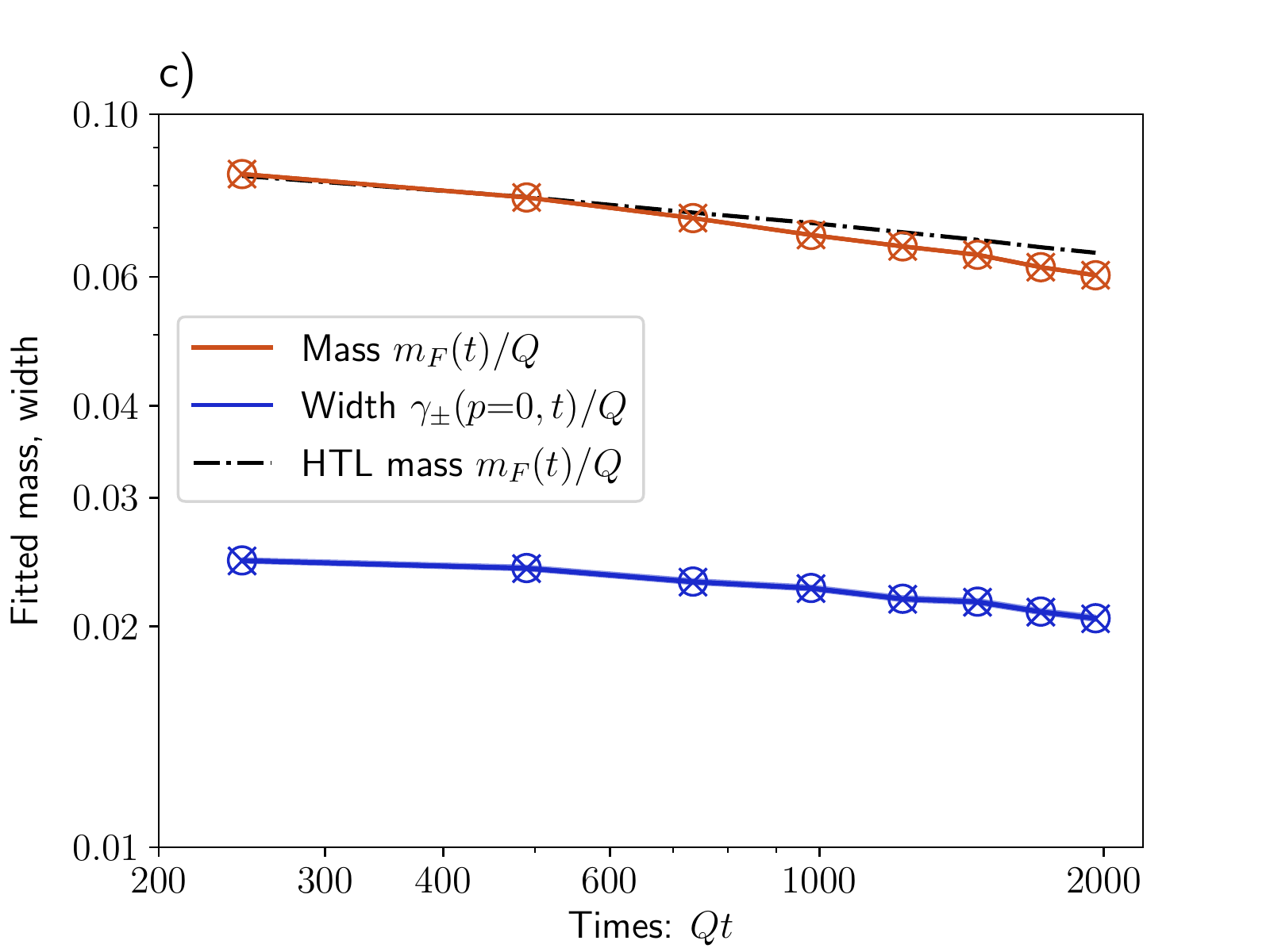}
\caption{
Spectral function $\rho_+ = \rho_V^0$ at $p=0$ for different times as a function of frequency, with all dimensionful quantities rescaled {\em a)} by $\Q$ and {\em b)} by $m_F(t)$.
{\em c)} 
\changeflag{Evolution of the mass $m_F(t)$ and width $\gamma(t,p{=}0)$ for the zero mode $p=0$ as function of time $\Q t$, shown on a log-log scale. Open circles / crosses correspond to extractions in the frequency / time domain using \eq\eqref{eq:rhoPlus_fit} / \eq\eqref{eq:lorD_dt}, respectively, which are in excellent agreement with each other.
The black dot-dashed line shows the HTL fermion mass using \eq\eqref{eq:mf_def}.}
}
\label{fig:p0_fit_dt}
\end{figure}

\subsection{Time evolution}
So far we have studied the behavior of the quark spectral function at a fixed reference time $\Q \tpert = 1500$, in  the self-similar evolution of a highly occupied gluon plasma. By focusing on the behavior of the quark spectral function at vanishing spatial momentum $p=0$, we will now investigate the non-equilibrium evolution of the quark screening mass and damping rate, where at different evolution times $Q\tpert$ different separations of hard and soft scales in the system can be accessed~\cite{Kurkela:2012hp,Berges:2013fga,York:2014wja,Mace:2016svc}.

We present our results for the zero momentum spectral function $\rho_{+}(t,\omega,p{=}0)$ in the top panel of \fig\ref{fig:p0_fit_dt}, where we show the frequency dependence of the spectral function at different times $Qt=245, 735, 1960$. When plotting all dimensionful quantities in terms of $\Q$, the qualitative features of the quark spectral function in \fig\ref{fig:p0_fit_dt}a) still remains essentially the same at all times. By expressing all dimensionful scales in units of the mass $m_F(t)$, this statement can be made quantitative, as shown in \fig\ref{fig:p0_fit_dt}b), where all curves fall on top of each other to good accuracy, indicating that $m_f(t)$ is the only relevant scale. 

The time dependence of the fermion mass $m_f(t) \equiv \omega_\pm(t,p{=}0)$ and damping rate $\gamma(t,p{=}0)$ are depicted in \fig\ref{fig:p0_fit_dt}c). 
We find that the time dependence of the effective quark mass exhibits an approximate $m_{f}(t)/\Q \propto (\Q\tpert)^{-1/7}$ scaling behavior, as can be expected by evaluating the perturbative expression in \eq\eqref{eq:mf_def} for the self-similar scaling behavior of the gluon distribution in \eq\eqref{eq:f_fp}. Direct comparison of the perturbative expression in \eq\eqref{eq:mf_def}, which is shown in terms of a black dashed line in \fig\ref{fig:p0_fit_dt}, indicates that the extracted value of $m_{f}(t)$ can be described rather accurately with deviations up to a $\lesssim 10$\% level. 

Similarly to the effective quark mass, the quark damping rate $\gamma(t,p{=}0)$ also decreases as a function of time, as visible in \fig\ref{fig:p0_fit_dt}c). More precisely, its time-evolution is approximately the same $\gamma(t,p{=}0) \sim m_f(t)$ in the plotted time range. 
This is in contrast to perturbative HTL expectations \cite{Braaten:1992gd,Blaizot:2001nr}, where the damping rate is expected to be proportional to the effective temperature that scales as $\gamma^\HTL(t,p{=}0) \propto g^2 T^*(t) \sim \Q (\Q t)^{-3/7}$ in the self-similar regime. The latter would imply that the associated damping rate would decrease more rapidly in time than the thermal mass, resulting in increasingly sharp quasi-particle peaks at late times. Such behavior has indeed been observed for the gluon spectral function in \re\cite{Boguslavski:2018beu}. In contrast, we find that due to the similar decrease of quark mass and damping rate, the spectral functions in \fig\ref{fig:p0_fit_dt}b) do not experience a significant sharpening of the quasi-particle peaks over the course of the evolution, contrary to  the perturbative expectation.

%% file: Conclusion.tex

\section{Conclusions and Outlook}
\label{sec:conc}
In this work, we have presented a novel method to perform non-perturbative real time calculations of fermion spectral functions in highly occupied plasmas. Based on a classical-statistical description of bosonic quantum fields, the fermion spectral function can be calculated by solving linearized evolution equations for fermions in the background of dynamical bosonic fields. 
Since only an individual momentum mode needs to be simulated at the same time, obtaining the spectral function is comutationally much less demanding than a full simulation of the fermion sector~\cite{Mace:2016shq,Mace:2019cqo}. 

Based on this approach, we studied the behavior of the quark spectral function in the vicinity of a so-called non-thermal fixed point where the non-equilibrium plasma exhibits a self-similar scaling behavior. We observe Landau damping and clear quasi-particle peaks for which we extracted dispersion relations, decay widths and residues as function of the momentum. Generally the dispersion relation and residues are well reproduced by leading order HTL perturbation theory, with a single parameter -- the quark screening mass $m_f$ -- which we extract consistently within the HTL framework. Beyond the familiar structures of leading order HTL perturbation theory, we find that the non-perturbative spectral functions also exhibit a finite decay width $\gamma^{+}(t,p)$, and we extracted its time and momentum dependence from our simulations. Unexpectedly, the damping rate of the zero momentum $\gamma^{+}(t,p{=}0)$ decreases much slower than in HTL perturbation theory and remains of the same order as the mass $\gamma^{+}(t,p{=}0) \sim m_f(t)$, a feature that has been observed also in lower dimensional gluon spectral functions \cite{Boguslavski:2021buh}.

Beyond the results presented in this paper, the methodology to perform non-perturbative calculations of fermion spectral functions provides an interesting new tool to benchmark and perhaps improve perturbative calculations  in the presence of strong gauge or scalar fields. Some possible extensions could include, e.g., the analysis of quark spectral functions in an expanding QCD plasma, or the investigation of the behavior of highly-energetic or heavy-flavor quarks, which we intend to pursue in the future.

\section*{Acknowledgements}

We would like to thank A.\ Kurkela, A.\ Pi\~{n}eiro Orioli, J.\ Peuron, S.~Sharma and L.~v.~Smekal for collaboration on related topics. 
This work has been supported by the European Research Council under grant no.\ ERC-2015-
CoG-681707, by the EU Horizon 2020 research and innovation programme, STRONG-2020
project (grant agreement No 824093) by the Academy of Finland, project 321840, by the Deutsche Forschungsgemeinschaft (DFG, German Research Foundation) through the CRC-TR 211
``Strong-interaction matter under extreme conditions'' Project number 315477589, \changeflag{and by the Austrian Science Fund (FWF) under project P 34455-N}.  We gratefully acknowledge the
National Energy Research Scientific Computing
Center, a U.S. Department of Energy Office of Science
User Facility supported under Contract No. DE-AC02-
05CH11231, 
the Vienna Scientific Cluster (VSC), Austria, and
CSC - IT Center for Science, Finland, for providing
computational resources. The content of this article does not reflect
the official opinion of the European Union and responsibility for the information and views
expressed therein lies entirely with the authors.

%% file: HTL.tex

\section{Spectral functions in HTL perturbation theory}
\label{sec:HTL}


We recall here the results of  perturbative calculations in the HTL framework that serve as a point of reference for interpreting our numerical results. We consider here the fermionic spectral function as computed within the hard-loop (HTL) framework at leading order. 
The fermionic HTL self-energy $\Sigma$ is known (see \cite{Laine:2016hma,Bellac:2011kqa}, and \cite{Mrowczynski:2000ed,Ghiglieri:2020dpq} in the case of a more general non-thermal state) and reads
\begin{align}
\label{eq:fermiSelfEn}
 \Sigma(\omega,\mbf p) \,&= m_f^2 \int \dfrac{\ud\Omega}{4\pi}\, \dfrac{\gamma^\mu \hat{K}_\mu}{P \cdot \hat{K}} 
\end{align}
with $\hat{K} = (1, \mbf k/k)$, $P = (\omega + i\epsilon, \mbf p)$ and metric signature $(1,-\mbf 1)$. 
The fermion mass $m_f$ is given by  
\begin{align}
 m_f^2 = \dfrac{(d-1)\, g^2 C_F}{4} \int \dfrac{\ud^d p}{(2\pi)^d}\, \frac{2f_g(p) + N_f(f_q(p) + \bar{f}_q(p))}{p} 
\end{align}
with $C_F = (N_c^2-1)/(2N_c)$, $N_f$ fermionic fields with distributions $f_q(p)$ (and $\bar{f}_q(p)$) for particles (anti-particles) and the distribution function of gauge fields $f_g(p)$. We are here working in the limit $f_g \gg 1$, where the fermionic contributions can be neglected and we can connect the formula to the asymptotic mass of gluons as
\begin{align} \label{eq:mf_def_app}
 m_f^2 \,&= \dfrac{(d-1)\, g^2 C_F}{2} \int \dfrac{\ud^d p}{(2\pi)^d}\, \dfrac{1}{p}\, f_g(p) \nonumber \\
 &= \dfrac{C_F}{2 N_c}\, m_g^2\,.
\end{align}
After evaluating the angular integration in \eq\eqref{eq:fermiSelfEn}, the dressed propagator can be written as
\begin{align}
\label{eq:Sinv}
 iS^{-1}(\omega,\mbf p) &= p_\mu \gamma^\mu - \Sigma(\omega,\mbf p) \nonumber \\
 &= A_0(\omega, p) \gamma^0 - A_V(\omega, p)\,\frac{p^j}{p} \gamma^j\,,
\end{align}
with functions
\begin{align}
 A_0(\omega, p) \,&= \omega - \dfrac{m_f^2}{p}\,Q_0(x) \nonumber \\
 A_V(\omega, p) \,&= p + \dfrac{m_f^2}{p} \left( 1 - x Q_0(x) \right),
\end{align}
for $x = \omega/p$ and the Legendre function
\begin{align}
 Q_0(x) \,&= \dfrac{1}{2}\, \ln \left( \dfrac{x+1}{x-1} \right) \nonumber \\
 \,&= \dfrac{1}{2}\, \ln \left| \dfrac{x+1}{x-1} \right| - \dfrac{i\pi}{2}\,\theta(1-x^2) \,.
\end{align}

The HTL propagator can be brought into the form 
\begin{align}
 \label{eq_rho_Lambda_decomp_app}
 S(\omega,\mbf p) = \dfrac{1}{2 E_p} \left( \Lambda_+(\mbf p)\,\Delta_+(\omega,p) + \Lambda_-(-\mbf p)\,\Delta_-(\omega,p) \right),
\end{align}
with the propagators
\begin{align}
\label{eq:DeltaPM}
 \Delta_{\pm}(\omega,p) &= \left( A_0(\omega, p) \mp A_V(\omega, p) \right)^{-1} \nonumber \\
 &= \left( \omega \mp p - \dfrac{m_f^2}{2p} \left[ \left( 1 \mp \dfrac{\omega}{p} \right) \ln\dfrac{\omega + p}{\omega - p} \pm 2 \right] \right)^{-1}.
\end{align}
This can be checked explicitly by multiplying $iS^{-1}(\omega,\mbf p) \times (-iS(\omega,\mbf p)) = 1$ using \eqref{eq:Sinv} and \eqref{eq_rho_Lambda_decomp_app}. It is important to note that $\Delta_\pm(\omega,p)$ are related by symmetry as
\begin{align}
    \mrm{Re}\,\Delta_\pm(\omega,p) &= -\mrm{Re}\,\Delta_\mp(-\omega,p) \nonumber \\
    \mrm{Im}\,\Delta_\pm(\omega,p) &= \mrm{Im}\,\Delta_\mp(-\omega,p).
\end{align}
Therefore, it is sufficient to restrict oneself to only $\Delta_+(\omega,p)$ since it already contains all the relevant information, or, alternatively, to consider both $\Delta_\pm$ but to restrict the frequency domain to positive values. In this paper we follow the former strategy, and only consider the particle components, but including both signs of $\omega$.
Thus we define the dispersion relations $\omega_\pm(p)$ as the poles of $\Delta_+(\omega,p)$ at $\omega = \omega_+(p)$ for $\omega > 0$ and $\omega = -\omega_-(p)$ for $\omega > 0$, i.e., by solving $A_0 + A_V = 0$ for both signs of the frequency. The symmetries imply that $\Delta_-$ will have a pole at $\omega = -\omega_+$ and another one at $\omega = \omega_-$. Note that the quasiparticle at $\omega_+$ corresponds to a particle-like state with positive helicity over chirality ratio $\chi = +1$ since it is the positive frequency pole of the function $\Delta_+$ multiplying the operator $\Lambda_+(\mbf p) \approx \gamma^0 p + \gamma^j p_j$, where we neglected a possible mass $m$. Likewise, the quasiparticle at $\omega_-$ is associated to an antiparticle-like ratio $\chi = -1$ due to its multiplication with $\Lambda_-(-\mbf p)$. 

To discuss the excitation spectrum in more detail, we compute the spectral function as the imaginary part of the propagators $\Delta_\pm(\omega,p)$ as
\begin{align} 
 \label{eq:HTL_rho_app}
 \rho_{\pm}(\omega,p) =&\, -2\, \mrm{Im}\, \Delta_{\pm}(\omega,p) \nonumber \\
 =&\, 2\pi \left[ Z_\pm(p)\delta(\omega - \omega_\pm(p)) + Z_\mp(p)\delta(\omega + \omega_\mp(p)) \right] \nonumber \\
 &\, + 2\pi\,\beta_\pm(\omega/p,p)
\end{align}
As commonly done, we distinguish here quasiparticle excitations that correspond to the delta-function peaks and a Landau damping part $\beta_\pm$ for $p>0$ given by
\begin{align}
 \label{eq:HTL_Landau}
 &\beta_\pm(x,p) \nonumber \\
 &=  \dfrac{m_f^2}{2p}\,(1\mp x) \theta(1-x^2)\left[\left( p(1\mp x)\pm \dfrac{m_f^2}{2p}\left[ (1\mp x) \right. \right. \right. \nonumber \\
 &~ \times \left.\left.\left. \ln \left| \dfrac{x+1}{x-1} \right| \pm 2 \right] \right)^2 + \dfrac{\pi^2 m_f^4}{4p^2}\,(1\mp x)^2 \right]^{-1}.
\end{align}
This region only exists at low frequencies $|\omega| < p$ and vanishes at $p=0$ identically, as can be seen in \eqref{eq:DeltaPM} since 
\begin{align}
    \Delta_{\pm}(\omega,p{=}0) = \frac{\omega}{\omega^2-m_f^2}
\end{align}
is real-valued with poles at $\omega_\pm(p{=}0) = m_f$ and residues $Z_\pm(p{=}0) = 1/2$. For $p=0$ the HTL retarded propagator has therefore the simple form $S(\omega,p{=}0) = \gamma^0 \omega/(\omega^2-m_f^2)$. 

For general momenta, the dispersion relations cannot be solved analytically. For small momenta $p \ll m_f$ they read
\begin{align}
 \omega_\pm(p) \simeq m_f \pm \dfrac{1}{3}\, p
\end{align}
and for large momenta $p \gg m_f$ 
\begin{align}
 \omega_+(p) \,&\simeq p + \dfrac{m_l^2}{2 p} \\
 \omega_-(p) \,&\simeq p + \dfrac{2p}{e} \exp\left( -\dfrac{2p^2}{m_f^2} \right),
\end{align}
where we have used the definition $m_l = \sqrt{2}\,m_f$ of the asymptotic fermion mass. In fact, the expression for $\omega_+(p)$ at large momenta can be interpreted as a large $p$ expansion of the relativistic dispersion relation $\omega_+(p) \simeq \sqrt{m_l^2 + p^2}$. 

The quasiparticle residues are given by 
\begin{align}
 Z_\pm(p) = \dfrac{\omega_\pm^2(p) - p^2}{2m_f^2}\,.
\end{align}
For low momenta $p \ll m_f$, this leads to 
\begin{align}
    Z_\pm(p) \simeq \frac{1}{2} \pm \frac{p}{3 m_f}
\end{align}
and for large momenta $p \gg m_f$ to
\begin{align}
    Z_+(p) &\simeq 1 - \frac{m_f^2}{2p^2}\left( \log\left(\frac{2p^2}{m_f^2}\right) - 1 \right) \\
    Z_-(p) &\simeq \frac{2p^2}{m_f^2}\, \exp \left( -\frac{2p^2}{m_f^2} -1 \right)
\end{align}
From canonical anticommutation relations, one obtains the important sum rule
\begin{align}
 1 \,&= \int_{-\infty}^\infty \dfrac{\ud \omega}{2\pi}\,\rho_{\pm}(\omega,p) \nonumber \\
 &= Z_+(p) + Z_-(p) + \int_{-1}^1 \ud x\,\beta_{\pm}(x,p)\,,
\end{align}
which is satisfied by the HTL spectral functions identically. Since $\beta_{\pm} \geq 0$ and $Z_\pm>0$, the residues are bound by unity from above. Numerically, the HTL result gives $0.8 \leq Z_+(p) + Z_-(p) \leq 1$ for all momenta \cite{Bellac:2011kqa}, implying that quasiparticle excitations always provide the dominant contributions to the spectral function. At low momenta $p \ll m_f$ both quasiparticle excitations have nearly equal residues around $1/2$. At high momenta $p \gg m_f$ the residue of the plasmino $Z_-(p)$ falls off  exponentially while the fermion with positive chirality survives with $Z_+(p) \approx 1$.

%% file: FermionsUneqTime.bbl
\providecommand{\href}[2]{#2}\begingroup\raggedright\begin{thebibliography}{10}

\bibitem{Lappi:2006fp}
T.~Lappi and L.~McLerran, {\it {Some features of the glasma}},
  \href{http://dx.doi.org/10.1016/j.nuclphysa.2006.04.001}{{\em Nucl. Phys. A}
  {\bf 772} (2006) 200} [\href{http://arXiv.org/abs/hep-ph/0602189}{{\tt
  arXiv:hep-ph/0602189}}].

\bibitem{Gelis:2015gza}
F.~Gelis, {\it {Initial state and thermalization in the Color Glass Condensate
  framework}},  \href{http://dx.doi.org/10.1142/S0218301315300088}{{\em Int. J.
  Mod. Phys. E} {\bf 24} (2015) 1530008}
  [\href{http://arXiv.org/abs/1508.07974}{{\tt arXiv:1508.07974 [hep-ph]}}].

\bibitem{Schlichting:2019abc}
S.~Schlichting and D.~Teaney, {\it The first fm/c of heavy-ion collisions},
  \href{http://dx.doi.org/10.1146/annurev-nucl-101918-023825}{{\em Ann. Rev.
  Nucl. Part. Sci.} {\bf 69} (2019) 447}
  [\href{http://arXiv.org/abs/1908.02113}{{\tt arXiv:1908.02113 [nucl-th]}}].

\bibitem{Berges:2020fwq}
J.~Berges, M.~P. Heller, A.~Mazeliauskas and R.~Venugopalan, {\it
  {Thermalization in QCD: theoretical approaches, phenomenological
  applications, and interdisciplinary connections}},
  \href{http://arXiv.org/abs/2005.12299}{{\tt arXiv:2005.12299 [hep-th]}}.

\bibitem{Allahverdi:2010xz}
R.~Allahverdi, R.~Brandenberger, F.-Y. Cyr-Racine and A.~Mazumdar, {\it
  Reheating in inflationary cosmology: Theory and applications},
  \href{http://dx.doi.org/10.1146/annurev.nucl.012809.104511}{{\em Ann. Rev.
  Nucl. Part. Sci.} {\bf 60} (2010) 27}
  [\href{http://arXiv.org/abs/1001.2600}{{\tt arXiv:1001.2600 [hep-th]}}].

\bibitem{Amin:2014eta}
M.~A. Amin, M.~P. Hertzberg, D.~I. Kaiser and J.~Karouby, {\it Nonperturbative
  dynamics of reheating after inflation: A review},
  \href{http://dx.doi.org/10.1142/S0218271815300037}{{\em Int. J. Mod. Phys. D}
  {\bf 24} (2014) 1530003} [\href{http://arXiv.org/abs/1410.3808}{{\tt
  arXiv:1410.3808 [hep-ph]}}].

\bibitem{Lozanov:2019jxc}
K.~D. Lozanov, {\it Lectures on reheating after inflation},
  \href{http://arXiv.org/abs/1907.04402}{{\tt arXiv:1907.04402 [astro-ph.CO]}}.

\bibitem{Ipp:2020nfu}
A.~Ipp, D.~I. M\"uller and D.~Schuh, {\it {Jet momentum broadening in the
  pre-equilibrium Glasma}},
  \href{http://dx.doi.org/10.1016/j.physletb.2020.135810}{{\em Phys. Lett. B}
  {\bf 810} (2020) 135810} [\href{http://arXiv.org/abs/2009.14206}{{\tt
  arXiv:2009.14206 [hep-ph]}}].

\bibitem{Martinez:2008di}
M.~Martinez and M.~Strickland, {\it {Pre-equilibrium dilepton production from
  an anisotropic quark-gluon plasma}},
  \href{http://dx.doi.org/10.1103/PhysRevC.78.034917}{{\em Phys. Rev. C} {\bf
  78} (2008) 034917} [\href{http://arXiv.org/abs/0805.4552}{{\tt
  arXiv:0805.4552 [hep-ph]}}].

\bibitem{Coquet:2021cuv}
M.~Coquet, X.~Du, J.-Y. Ollitrault, S.~Schlichting and M.~Winn, {\it
  {Intermediate mass dileptons as pre-equilibrium probes in heavy ion
  collisions}},  \href{http://arXiv.org/abs/2104.07622}{{\tt arXiv:2104.07622
  [nucl-th]}}.

\bibitem{Jeon:2004dh}
S.~Jeon, {\it {The Boltzmann equation in classical and quantum field theory}},
  \href{http://dx.doi.org/10.1103/PhysRevC.72.014907}{{\em Phys. Rev. C} {\bf
  72} (2005) 014907} [\href{http://arXiv.org/abs/hep-ph/0412121}{{\tt
  arXiv:hep-ph/0412121}}].

\bibitem{Mueller:2002gd}
A.~H. Mueller and D.~T. Son, {\it On the equivalence between the {Boltzmann}
  equation and classical field theory at large occupation numbers},
  \href{http://dx.doi.org/10.1016/j.physletb.2003.12.047}{{\em Phys. Lett. B}
  {\bf 582} (2004) 279} [\href{http://arXiv.org/abs/hep-ph/0212198}{{\tt
  arXiv:hep-ph/0212198}}].

\bibitem{Berges:2004yj}
J.~Berges, {\it {Introduction to nonequilibrium quantum field theory}},
  \href{http://dx.doi.org/10.1063/1.1843591}{{\em AIP Conf. Proc.} {\bf 739}
  (2004) 3} [\href{http://arXiv.org/abs/hep-ph/0409233}{{\tt
  arXiv:hep-ph/0409233}}].

\bibitem{Berges:2013fga}
J.~Berges, K.~Boguslavski, S.~Schlichting and R.~Venugopalan, {\it {Universal
  attractor in a highly occupied non-Abelian plasma}},
  \href{http://dx.doi.org/10.1103/PhysRevD.89.114007}{{\em Phys. Rev. D} {\bf
  89} (2014) 114007} [\href{http://arXiv.org/abs/1311.3005}{{\tt
  arXiv:1311.3005 [hep-ph]}}].

\bibitem{Berges:2013eia}
J.~Berges, K.~Boguslavski, S.~Schlichting and R.~Venugopalan, {\it {Turbulent
  thermalization process in heavy-ion collisions at ultrarelativistic
  energies}},  \href{http://dx.doi.org/10.1103/PhysRevD.89.074011}{{\em Phys.
  Rev. D} {\bf 89} (2014) 074011} [\href{http://arXiv.org/abs/1303.5650}{{\tt
  arXiv:1303.5650 [hep-ph]}}].

\bibitem{Figueroa:2020rrl}
D.~G. Figueroa, A.~Florio, F.~Torrenti and W.~Valkenburg, {\it {The art of
  simulating the early Universe -- Part I}},
  \href{http://arXiv.org/abs/2006.15122}{{\tt arXiv:2006.15122 [astro-ph.CO]}}.

\bibitem{Boguslavski:2018beu}
K.~Boguslavski, A.~Kurkela, T.~Lappi and J.~Peuron, {\it Spectral function for
  overoccupied gluodynamics from real-time lattice simulations},
  \href{http://dx.doi.org/10.1103/PhysRevD.98.014006}{{\em Phys. Rev.} {\bf
  D98} (2018) 014006} [\href{http://arXiv.org/abs/1804.01966}{{\tt
  arXiv:1804.01966 [hep-ph]}}].

\bibitem{Boguslavski:2021buh}
K.~Boguslavski, A.~Kurkela, T.~Lappi and J.~Peuron, {\it Broad excitations in a
  2+1d overoccupied gluon plasma},
  \href{http://dx.doi.org/10.1007/JHEP05(2021)225}{{\em JHEP} {\bf 05} (2021)
  225} [\href{http://arXiv.org/abs/2101.02715}{{\tt arXiv:2101.02715
  [hep-ph]}}].

\bibitem{Mace:2016shq}
M.~Mace, N.~Mueller, S.~Schlichting and S.~Sharma, {\it {Non-equilibrium study
  of the Chiral Magnetic Effect from real-time simulations with dynamical
  fermions}},  \href{http://dx.doi.org/10.1103/PhysRevD.95.036023}{{\em Phys.
  Rev. D} {\bf 95} (2017) 036023} [\href{http://arXiv.org/abs/1612.02477}{{\tt
  arXiv:1612.02477 [hep-lat]}}].

\bibitem{Mace:2019cqo}
M.~Mace, N.~Mueller, S.~Schlichting and S.~Sharma, {\it Chiral instabilities
  and the onset of chiral turbulence in {QED} plasmas},
  \href{http://dx.doi.org/10.1103/PhysRevLett.124.191604}{{\em Phys. Rev.
  Lett.} {\bf 124} (2020) 191604} [\href{http://arXiv.org/abs/1910.01654}{{\tt
  arXiv:1910.01654 [hep-ph]}}].

\bibitem{Tripolt:2020irx}
R.-A. Tripolt, D.~H. Rischke, L.~von Smekal and J.~Wambach, {\it {Fermionic
  excitations at finite temperature and density}},
  \href{http://dx.doi.org/10.1103/PhysRevD.101.094010}{{\em Phys. Rev. D} {\bf
  101} (2020) 094010} [\href{http://arXiv.org/abs/2003.11871}{{\tt
  arXiv:2003.11871 [hep-ph]}}].

\bibitem{Epelbaum:2011pc}
T.~Epelbaum and F.~Gelis, {\it {Role of quantum fluctuations in a system with
  strong fields: Spectral properties and Thermalization}},
  \href{http://dx.doi.org/10.1016/j.nuclphysa.2011.09.019}{{\em Nucl. Phys. A}
  {\bf 872} (2011) 210} [\href{http://arXiv.org/abs/1107.0668}{{\tt
  arXiv:1107.0668 [hep-ph]}}].

\bibitem{PineiroOrioli:2018hst}
A.~Pi{\~n}eiro~Orioli and J.~Berges, {\it {Breaking the fluctuation-dissipation
  relation by universal transport processes}},
  \href{http://dx.doi.org/10.1103/PhysRevLett.122.150401}{{\em Phys. Rev.
  Lett.} {\bf 122} (2019) 150401} [\href{http://arXiv.org/abs/1810.12392}{{\tt
  arXiv:1810.12392 [cond-mat.quant-gas]}}].

\bibitem{Boguslavski:2019ecc}
K.~Boguslavski and A.~Pi{\~n}eiro~Orioli, {\it Unraveling the nature of
  universal dynamics in {$O(N)$} theories},
  \href{http://dx.doi.org/10.1103/PhysRevD.101.091902}{{\em Phys. Rev. D} {\bf
  101} (2020) 091902} [\href{http://arXiv.org/abs/1911.04506}{{\tt
  arXiv:1911.04506 [hep-ph]}}].

\bibitem{Aarts:2001yx}
G.~Aarts, {\it {Spectral function at high temperature in the classical
  approximation}},  \href{http://dx.doi.org/10.1016/S0370-2693(01)01081-4}{{\em
  Phys. Lett. B} {\bf 518} (2001) 315}
  [\href{http://arXiv.org/abs/hep-ph/0108125}{{\tt arXiv:hep-ph/0108125}}].

\bibitem{Schlichting:2019tbr}
S.~Schlichting, D.~Smith and L.~von Smekal, {\it {Spectral functions and
  critical dynamics of the $O(4)$ model from classical-statistical lattice
  simulations}},  \href{http://dx.doi.org/10.1016/j.nuclphysb.2019.114868}{{\em
  Nucl. Phys. B} {\bf 950} (2020) 114868}
  [\href{http://arXiv.org/abs/1908.00912}{{\tt arXiv:1908.00912 [hep-lat]}}].

\bibitem{Schweitzer:2020noq}
D.~Schweitzer, S.~Schlichting and L.~von Smekal, {\it {Spectral functions and
  dynamic critical behavior of relativistic $Z_2$ theories}},
  \href{http://dx.doi.org/10.1016/j.nuclphysb.2020.115165}{{\em Nucl. Phys. B}
  {\bf 960} (2020) 115165} [\href{http://arXiv.org/abs/2007.03374}{{\tt
  arXiv:2007.03374 [hep-lat]}}].

\bibitem{Braaten:1992gd}
E.~Braaten and R.~D. Pisarski, {\it {Calculation of the quark damping rate in
  hot QCD}},  \href{http://dx.doi.org/10.1103/PhysRevD.46.1829}{{\em Phys. Rev.
  D} {\bf 46} (1992) 1829}.

\bibitem{Blaizot:2001nr}
J.-P. Blaizot and E.~Iancu, {\it {The Quark gluon plasma: Collective dynamics
  and hard thermal loops}},
  \href{http://dx.doi.org/10.1016/S0370-1573(01)00061-8}{{\em Phys. Rept.} {\bf
  359} (2002) 355} [\href{http://arXiv.org/abs/hep-ph/0101103}{{\tt
  arXiv:hep-ph/0101103}}].

\bibitem{Mrowczynski:2000ed}
S.~Mrowczynski and M.~H. Thoma, {\it {Hard loop approach to anisotropic
  systems}},  \href{http://dx.doi.org/10.1103/PhysRevD.62.036011}{{\em Phys.
  Rev.} {\bf D62} (2000) 036011}
  [\href{http://arXiv.org/abs/hep-ph/0001164}{{\tt arXiv:hep-ph/0001164
  [hep-ph]}}].

\bibitem{Bellac:2011kqa}
M.~L. Bellac, {\em {Thermal Field Theory}}.
\newblock Cambridge Monographs on Mathematical Physics. Cambridge University
  Press, 2011.

\bibitem{Baier:1991dy}
R.~Baier, G.~Kunstatter and D.~Schiff, {\it {High temperature fermion
  propagator: Resummation and gauge dependence of the damping rate}},
  \href{http://dx.doi.org/10.1103/PhysRevD.45.R4381}{{\em Phys. Rev. D} {\bf
  45} (1992) R4381}.

\bibitem{Rebhan:1992ak}
A.~Rebhan, {\it {Comment on `high temperature fermion propagator: Resummation
  and gauge dependence of the damping rate'}},
  \href{http://dx.doi.org/10.1103/PhysRevD.46.4779}{{\em Phys. Rev. D} {\bf 46}
  (1992) 4779} [\href{http://arXiv.org/abs/hep-ph/9204210}{{\tt
  arXiv:hep-ph/9204210}}].

\bibitem{Aarts:1998td}
G.~Aarts and J.~Smit, {\it {Real time dynamics with fermions on a lattice}},
  \href{http://dx.doi.org/10.1016/S0550-3213(99)00320-X}{{\em Nucl. Phys. B}
  {\bf 555} (1999) 355} [\href{http://arXiv.org/abs/hep-ph/9812413}{{\tt
  arXiv:hep-ph/9812413}}].

\bibitem{Kasper:2014uaa}
V.~Kasper, F.~Hebenstreit and J.~Berges, {\it {Fermion production from
  real-time lattice gauge theory in the classical-statistical regime}},
  \href{http://dx.doi.org/10.1103/PhysRevD.90.025016}{{\em Phys. Rev. D} {\bf
  90} (2014) 025016} [\href{http://arXiv.org/abs/1403.4849}{{\tt
  arXiv:1403.4849 [hep-ph]}}].

\bibitem{Kurkela:2016mhu}
A.~Kurkela, T.~Lappi and J.~Peuron, {\it {Time evolution of linearized gauge
  field fluctuations on a real-time lattice}},
  \href{http://dx.doi.org/10.1140/epjc/s10052-016-4523-9}{{\em Eur. Phys. J. C}
  {\bf 76} (2016) 688} [\href{http://arXiv.org/abs/1610.01355}{{\tt
  arXiv:1610.01355 [hep-lat]}}].

\bibitem{Berges:2008mr}
J.~Berges, S.~Scheffler and D.~Sexty, {\it {Turbulence in nonabelian gauge
  theory}},  \href{http://dx.doi.org/10.1016/j.physletb.2009.10.032}{{\em Phys.
  Lett. B} {\bf 681} (2009) 362} [\href{http://arXiv.org/abs/0811.4293}{{\tt
  arXiv:0811.4293 [hep-ph]}}].

\bibitem{Kurkela:2011ti}
A.~Kurkela and G.~D. Moore, {\it {Thermalization in Weakly Coupled Nonabelian
  Plasmas}},  \href{http://dx.doi.org/10.1007/JHEP12(2011)044}{{\em JHEP} {\bf
  12} (2011) 044} [\href{http://arXiv.org/abs/1107.5050}{{\tt arXiv:1107.5050
  [hep-ph]}}].

\bibitem{Kurkela:2012hp}
A.~Kurkela and G.~D. Moore, {\it {UV Cascade in Classical Yang-Mills Theory}},
  \href{http://dx.doi.org/10.1103/PhysRevD.86.056008}{{\em Phys. Rev. D} {\bf
  86} (2012) 056008} [\href{http://arXiv.org/abs/1207.1663}{{\tt
  arXiv:1207.1663 [hep-ph]}}].

\bibitem{Berges:2012ev}
J.~Berges, S.~Schlichting and D.~Sexty, {\it {Over-populated gauge fields on
  the lattice}},  \href{http://dx.doi.org/10.1103/PhysRevD.86.074006}{{\em
  Phys. Rev. D} {\bf 86} (2012) 074006}
  [\href{http://arXiv.org/abs/1203.4646}{{\tt arXiv:1203.4646 [hep-ph]}}].

\bibitem{Schlichting:2012es}
S.~Schlichting, {\it {Turbulent thermalization of weakly coupled non-abelian
  plasmas}},  \href{http://dx.doi.org/10.1103/PhysRevD.86.065008}{{\em Phys.
  Rev. D} {\bf 86} (2012) 065008} [\href{http://arXiv.org/abs/1207.1450}{{\tt
  arXiv:1207.1450 [hep-ph]}}].

\bibitem{York:2014wja}
M.~C. Abraao~York, A.~Kurkela, E.~Lu and G.~D. Moore, {\it {UV cascade in
  classical Yang-Mills theory via kinetic theory}},
  \href{http://dx.doi.org/10.1103/PhysRevD.89.074036}{{\em Phys. Rev. D} {\bf
  89} (2014) 074036} [\href{http://arXiv.org/abs/1401.3751}{{\tt
  arXiv:1401.3751 [hep-ph]}}].

\bibitem{Mace:2016svc}
M.~Mace, S.~Schlichting and R.~Venugopalan, {\it {Off-equilibrium sphaleron
  transitions in the Glasma}},
  \href{http://dx.doi.org/10.1103/PhysRevD.93.074036}{{\em Phys. Rev. D} {\bf
  93} (2016) 074036} [\href{http://arXiv.org/abs/1601.07342}{{\tt
  arXiv:1601.07342 [hep-ph]}}].

\bibitem{Micha:2004bv}
R.~Micha and I.~I. Tkachev, {\it {Turbulent thermalization}},
  \href{http://dx.doi.org/10.1103/PhysRevD.70.043538}{{\em Phys. Rev. D} {\bf
  70} (2004) 043538} [\href{http://arXiv.org/abs/hep-ph/0403101}{{\tt
  arXiv:hep-ph/0403101}}].

\bibitem{Berges:2013lsa}
J.~Berges, K.~Boguslavski, S.~Schlichting and R.~Venugopalan, {\it Basin of
  attraction for turbulent thermalization and the range of validity of
  classical-statistical simulations},
  \href{http://dx.doi.org/10.1007/JHEP05(2014)054}{{\em JHEP} {\bf 05} (2014)
  054} [\href{http://arXiv.org/abs/1312.5216}{{\tt arXiv:1312.5216 [hep-ph]}}].

\bibitem{Boguslavski:2019fsb}
K.~Boguslavski, A.~Kurkela, T.~Lappi and J.~Peuron, {\it Highly occupied gauge
  theories in 2+1 dimensions: A self-similar attractor},
  \href{http://dx.doi.org/10.1103/PhysRevD.100.094022}{{\em Phys. Rev. D} {\bf
  100} (2019) 094022} [\href{http://arXiv.org/abs/1907.05892}{{\tt
  arXiv:1907.05892 [hep-ph]}}].

\bibitem{Braaten:1990it}
E.~Braaten and R.~D. Pisarski, {\it {Calculation of the gluon damping rate in
  hot QCD}},  \href{http://dx.doi.org/10.1103/PhysRevD.42.2156}{{\em Phys. Rev.
  D} {\bf 42} (1990) 2156}.

\bibitem{Laine:2016hma}
M.~Laine and A.~Vuorinen, {\it Basics of thermal field theory},
  \href{http://dx.doi.org/10.1007/978-3-319-31933-9}{{\em Lect. Notes Phys.}
  {\bf 925} (2016) pp.1} [\href{http://arXiv.org/abs/1701.01554}{{\tt
  arXiv:1701.01554 [hep-ph]}}].

\bibitem{Ghiglieri:2020dpq}
J.~Ghiglieri, A.~Kurkela, M.~Strickland and A.~Vuorinen, {\it Perturbative
  thermal {QCD:} formalism and applications},
  \href{http://dx.doi.org/10.1016/j.physrep.2020.07.004}{{\em Phys. Rept.} {\bf
  880} (2020) 1} [\href{http://arXiv.org/abs/2002.10188}{{\tt arXiv:2002.10188
  [hep-ph]}}].

\end{thebibliography}\endgroup
